\documentclass[aps,pra,reprint,superscriptaddress,nofootinbib,longbibliography,floatfix]{revtex4-2}

\usepackage{amsmath,amssymb,mathtools,bm}
\usepackage{graphicx}
\usepackage{placeins}
\usepackage{booktabs}
\usepackage{siunitx}
\usepackage{xcolor}
\usepackage{hyperref}

\definecolor{linkgray}{rgb}{0.18,0.25,0.27}
\hypersetup{
  colorlinks=true,
  linkcolor=linkgray,
  citecolor=linkgray,
  urlcolor=linkgray,
  pdftitle={Boundary-accessible work in a collision-charged topological quantum battery},
  pdfauthor={Osman Orcun Keskin; Ozgur E. Mustecaplioglu}
}
\graphicspath{{figures/}}
\newcommand{\W}{\mathcal{W}}
\newcommand{\Tr}{\operatorname{Tr}}

\begin{document}
\raggedbottom

\title{Boundary-accessible work in a collision-charged topological quantum battery}

\author{Osman Or\c{c}un Keskin}
\affiliation{Department of Physics, Ko\c{c} University, 34450 Sar\i yer, Istanbul, T\"urkiye}
\author{\"Ozg\"ur E. M\"ustecapl{\i}o\u{g}lu}
\affiliation{Department of Physics, Ko\c{c} University, 34450 Sar\i yer, Istanbul, T\"urkiye}
\affiliation{T\"UB\.ITAK Research Institute for Fundamental Sciences (TBAE), 41470 Gebze, Kocaeli, T\"urkiye}

\date{\today}

\begin{abstract}
Quantum batteries store energy in nonequilibrium quantum states. The ergotropy sets the work extractable by an arbitrary global unitary, but a device usually exchanges energy through a single local port. It remains unclear how much stored work such a port can deliver, and whether topological edge modes help. We study a dimerized spin chain of Su--Schrieffer--Heeger (SSH) type. Excited two-level ancillas charge it at one end by repeated collisions, and an initially empty two-level load discharges it at the same site. An exact free-fermion mapping yields a bound on the load work for all number-conserving quadratic extraction protocols. The load gains extractable work only if a single-particle mode of the battery holds more than half an excitation. Resonant charging fills the topological edge mode above this threshold. A single boundary pulse then delivers $98.6\%$ of the maximal load work allowed by the bound for representative parameters. Under the same six-collision controls, the trivial-phase load remains passive. Within the protocol that selects the load pulse for maximum delivered work, cycle power peaks at a finite collision number while delivered work continues to increase over the scanned range. Over the tested range, the advantage survives hopping disorder, which preserves chiral symmetry. On-site disorder breaks this symmetry and leaves a finite fraction of loads passive. Topology thus supplies a localized access channel, while spectral matching and symmetry decide whether stored energy becomes usable local work.
\end{abstract}

\maketitle

\section{Introduction}

Quantum batteries store energy in nonequilibrium quantum states. Stored energy and extractable work are distinct resources \cite{Allahverdyan2004,Vinjanampathy2016,Campaioli2024}; coherence can contribute to ergotropy \cite{Francica2020}, and correlations can alter its transport \cite{Simon2025}. Recent work emphasizes both full charge--store--discharge operation and scalable hardware: a field-wide perspective identifies the central experimental challenges \cite{Ferraro2026Perspective}, a superconducting processor demonstrates a many-cell charging advantage \cite{Hu2026Charging}, and giant-atom waveguides offer interference-protected energy transfer \cite{Yan2026Giant}. The ergotropy $\W$ is the maximum work extractable by a cyclic unitary on the complete battery. A laboratory port generally accesses a restricted part of that resource. We ask when energy stored in a structured battery becomes work accessible through a local port. We study this question in a hard-core Su--Schrieffer--Heeger (SSH) chain \cite{Su1979}, a minimal topological realization of the setting.

In the SSH chain, alternating hoppings support exponentially localized midgap boundary modes in the topological phase \cite{Ryu2002,Asboth2016}. Earlier SSH-battery work established phase-dependent charging, capacity, and global ergotropy \cite{Zhao2022}, while recent topological-battery proposals use photonic environments, critical dynamics, controlled edge-state transfer, and non-Hermitian SSH settings \cite{Lu2025,Kheiri2026,Yang2026,Zhou2026}. Repeated-interaction batteries exhibit collective and coherence-assisted charging effects \cite{Seah2021,Salvia2023}. Work transfer from a spin-chain battery to an external spin has also been studied in a topologically frustrated Ising chain \cite{Catalano2024Frustrating}. Here we consider a different mechanism: repeated boundary collisions populate an SSH edge orbital, and an initially empty load discharges the battery through the same site. We derive a tight load-population bound for number-conserving quadratic extraction and determine how closely a single boundary pulse approaches it. Cyclic-battery and measurement-assisted protocols address work extraction in other architectures \cite{Razzoli2025,Francica2017,Elyasi2025}.

Our main result links topology, occupation inversion, and local work. Resonant collisions fill the SSH edge mode beyond half occupation. A one-body bound fixes the maximal load work, and a single boundary pulse nearly reaches it. A sweep through the SSH transition at fixed controls separates these steps. We also quantify switching work, cycle power, extraction noise, and disorder. The Supplemental Material converts the parameters to superconducting-qubit timescales \cite{Liu2025Processor,Supplemental}.

\section{Model and work measures}

\subsection{Collision-charged SSH battery}

\begin{figure*}[!t]
  \centering
  \includegraphics{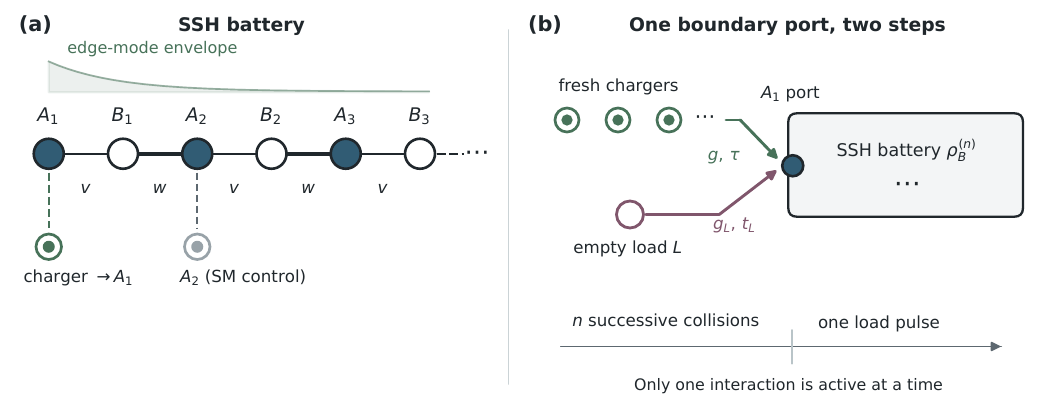}
  \caption{Model and protocol. (a) Hard-core SSH battery with intracell hopping $v$ and intercell hopping $w$. Filled (open) circles denote $A$ ($B$) sites; thick bonds mark the stronger hopping $w>v$ of the topological phase. The shaded curve sketches the left-edge probability $|\alpha_j|^2\propto(v/w)^{2(j-1)}$ [Eq.~\eqref{eq:edge}]. Chargers couple to the boundary site $A_1$; coupling to the interior site $A_2$ serves only as a control in the Supplemental Material~\cite{Supplemental}. (b) Protocol. Fresh excited chargers (ringed circles) collide in sequence with $A_1$, each with coupling $g$ for a duration $\tau$. After $n$ collisions, an initially empty two-level load $L$ (open circle) couples to the same site with coupling $g_L$ for a single pulse of duration $t_L$. Only one interaction is active at a time.}
  \label{fig:model}
\end{figure*}

The battery contains $N$ unit cells, or $2N$ hard-core-boson modes (equivalently, spin-$1/2$ sites) ordered as $A_1,B_1,\ldots,A_N,B_N$. Figure~\ref{fig:model} summarizes the site ordering and the charging--extraction protocol. With open boundaries,
\begin{align}
H_B={}&\epsilon_0\sum_{j=1}^{N}(n_{A_j}+n_{B_j})
+v\sum_{j=1}^{N}(\sigma_{A_j}^{+}\sigma_{B_j}^{-}+\mathrm{H.c.})\nonumber\\
&+w\sum_{j=1}^{N-1}(\sigma_{B_j}^{+}\sigma_{A_{j+1}}^{-}+\mathrm{H.c.}).
\label{eq:HB}
\end{align}
We write $\widehat N_B=\sum_{j=1}^{N}(n_{A_j}+n_{B_j})$ and $N_B=\Tr(\rho_B\widehat N_B)$ for the total battery excitation. We set $\hbar=1$ and express energies in units of the hopping scale $J$, defined by $v+w=2J$, and times in units of $J^{-1}$; powers are consequently quoted in $J^2/\hbar$. The uniform on-site term is $\epsilon_0=2.5J$, so all one-particle energies are positive and the empty chain is the unique many-body ground state. We compare $(v,w)=(1.5,0.5)J$ (trivial) with $(0.5,1.5)J$ (topological). For this termination, the topological left-edge profile obeys
\begin{equation}
\alpha_j=\alpha_1(-v/w)^{j-1},\qquad \xi^{-1}=\ln|w/v|,
\label{eq:edge}
\end{equation}
where the localization length $\xi$ is measured in unit cells. The Supplemental Material shows the finite-chain spectrum and edge profile \cite{Supplemental}.

A two-level charger with $H_A=\omega_A n_A$ interacts with site $\ell$ through $H_{\mathrm{int}}^{(\ell)}=g(\sigma_A^+\sigma_\ell^-+\mathrm{H.c.})$. In the Jordan--Wigner ordering the charger, and later the load, occupies position $0$ adjacent to $A_1$; the boundary couplings are therefore quadratic without an intervening parity string. This repeated-interaction construction follows the standard collision-model setting \cite{Ciccarello2022}. One collision applies
\begin{equation}
U_\ell=e^{-i\tau(H_A+H_B+H_{\mathrm{int}}^{(\ell)})},\quad
\rho'_B=\Tr_A[U_\ell(\lvert1\rangle\!\langle1\rvert\otimes\rho_B)U_\ell^\dagger].
\label{eq:collision}
\end{equation}
Unless stated otherwise, $g=0.3J$, $\tau=2J^{-1}$, and six fresh excited chargers address $A_1$. The baseline detuning is $\Delta_A=\omega_A-\epsilon_0=J$; it coincides with the lower edge of the upper bulk band in both phases. The edge-resonant setting is $\Delta_A=0$.

\subsection{Energy, ergotropy, and one-port extraction}
\label{sec:work-measures}

The stored battery energy is $E_B=\Tr(\rho_BH_B)$ because the ground energy is zero. If $r_k$ are the eigenvalues of $\rho_B$ in decreasing order and $\varepsilon_k$ those of $H_B$ in increasing order, the passive energy and global ergotropy are \cite{Allahverdyan2004}
\begin{equation}
E_{\mathrm{pas}}=\sum_k r_k\varepsilon_k,\qquad \W=E_B-E_{\mathrm{pas}}.
\label{eq:ergotropy}
\end{equation}
For number-conserving quadratic evolution in the Jordan--Wigner fermionic modes, define the one-body correlation matrix $(C_B)_{ij}=\Tr(\rho_Bc_j^\dagger c_i)$. Its eigenvectors are natural orbitals and its eigenvalues $\nu_j$ are natural occupations. Number-conserving quadratic unitaries act as $C_B\mapsto uC_Bu^\dagger$, while a quadratic battery Hamiltonian obeys $E_B=\Tr(hC_B)$. The trace inequality $\min_u\Tr(h\,uC_Bu^\dagger)=\sum_j\nu_j^\downarrow e_j^\uparrow$ therefore gives the quadratic ergotropy
\begin{equation}
\W^{(2)}=E_B-\sum_j\nu_j^{\downarrow}e_j^{\uparrow},
\label{eq:qerg}
\end{equation}
for any battery state, where $e_j$ are the one-particle energies. A related passivity construction for bosonic Gaussian operations is given in Ref.~\cite{Brown2016}. This definition distinguishes the unrestricted global resource from the work accessible to one-body mode rotations.

For local extraction, an empty two-level load with $H_L=\omega_Ln_L$ is coupled only to $A_1$,
\begin{align}
H_{BL}&=g_L(\sigma_{A_1}^{+}\sigma_L^{-}+\mathrm{H.c.}),\nonumber\\
U_{BL}&=e^{-it_L(H_B+H_L+H_{BL})}.
\label{eq:load}
\end{align}
The charged battery is number-diagonal, $[\rho_B,\widehat N_B]=0$; the load begins in $|0\rangle_L$, and $[U_{BL},\widehat N_B+n_L]=0$. The final joint state is therefore number-diagonal, so its one-mode load marginal commutes with $n_L$ and has the form $\rho_L=(1-p_L)|0\rangle\langle0|+p_L|1\rangle\langle1|$. Consequently, $E_L=\omega_Lp_L$ and
\begin{equation}
\W_L=\omega_L\max(0,2p_L-1).
\label{eq:loadwork}
\end{equation}
The load has positive ergotropy exactly when $p_L>1/2$. The boundary pulse transfers energy to the load and leaves an amount $\W_L$ available for subsequent local work extraction. For the diagonal load state considered here, this work can be extracted after decoupling by a cyclic local drive that interchanges the ground- and excited-state populations. We use ``delivered load work'' to denote this available ergotropy.

Any number-conserving quadratic extraction acts on the mode operators through a one-body unitary $u$ on the joint battery--load mode space. The initial correlation matrix is $0_L\oplus C_B$. The final load population is therefore $p_L=\bm u_B^\dagger C_B\bm u_B$, with $\bm u_B=P_Bu^\dagger|L\rangle$ and $\|\bm u_B\|\leq1$. The Rayleigh--Ritz theorem then gives
\begin{equation}
p_L\leq\nu_{\max}(C_B),\qquad
\W_L\leq\omega_L\max[0,2\nu_{\max}(C_B)-1].
\label{eq:bound}
\end{equation}
We call $\nu_{\max}>1/2$ a natural-orbital inversion. The population bound depends only on $C_B$; the work bound and activation criterion hold for Gaussian and non-Gaussian number-diagonal battery states. It is tight: the quadratic unitary that swaps the dominant natural orbital into the empty load attains $p_L=\nu_{\max}$. Within this class, natural-orbital inversion is therefore necessary and sufficient for load activation. Interacting number-conserving protocols can activate the load even when $\nu_{\max}\leq1/2$. A fixed boundary pulse approaches the bound to the extent that it overlaps the dominant orbital in space and energy.

For $n$ charging collisions, the aggregate charger switching work is $W_{\mathrm{sw}}^{(\mathrm{ch})}=\sum_{m=1}^{n}(\Delta E_B^{(m)}+\Delta E_A^{(m)})$ \cite{Strasberg2017}. For a vacuum battery and fresh excited chargers, the injected ancilla energy is $E_{\mathrm{in}}=-\sum_{m=1}^{n}\Delta E_A^{(m)}=\omega_A N_B$. Disconnecting the load contributes
\begin{equation}
W_{\mathrm{sw}}^{(L)}=\Delta E_B+E_L=-\langle H_{BL}\rangle_{t_L}.
\label{eq:switch}
\end{equation}
Positive switching work is supplied by the external controller; negative switching work is not credited. With aggregate overhead time $t_{\mathrm{oh}}$, including reset, dead time, and any prescribed duration of the final load-work extraction operation,
\begin{align}
\W_{\mathrm{net}}={}&\max\{0,\W_L-[W_{\mathrm{sw}}^{(\mathrm{ch})}]_+
                         -[W_{\mathrm{sw}}^{(L)}]_+\},\nonumber\\
P_{L,\mathrm{net}}={}&\frac{\W_{\mathrm{net}}}{n\tau+t_L+t_{\mathrm{oh}}},\nonumber\\
\eta_{\mathrm{cyc}}={}&\frac{\W_L}{E_{\mathrm{in}}+[W_{\mathrm{sw}}^{(\mathrm{ch})}]_+
                         +[W_{\mathrm{sw}}^{(L)}]_+}.
\label{eq:cycle}
\end{align}
Here $[x]_+=\max(x,0)$. The curves with $t_{\mathrm{oh}}=0$ use the ideal limit of negligible overhead, including negligible duration of the final local work-extraction operation. Finite overhead is represented by $t_{\mathrm{oh}}$; the local drive is assumed to attain the load ergotropy defined in Eq.~\eqref{eq:loadwork}. The reset is complete relaxation to vacuum with a prescribed dead time and no work credit. We call $\eta_{\mathrm{cyc}}$ the cycle resource yield. Energy balance gives $E_L\leq E_{\mathrm{in}}+[W_{\mathrm{sw}}^{(\mathrm{ch})}]_+ +[W_{\mathrm{sw}}^{(L)}]_+$; since $\W_L\leq E_L$, one has $\eta_{\mathrm{cyc}}\leq1$.

At resonance, $H-\epsilon_0\widehat N_{\mathrm{tot}}$ is a real nearest-neighbor hopping Hamiltonian on a bipartite chain. The charger and load are counted as sites on the sublattice opposite to $A_1$. The one-body propagator has the form $u=DOD^\dagger$, where $O$ is real orthogonal and $D=\Gamma^{1/2}$ for the chiral operator $\Gamma=\operatorname{diag}(I_A,-I_B)$; equivalently, $D_{jj}=1$ on the $A$ sublattice and $D_{jj}=i$ on the $B$ sublattice. Hence $C_B=DRD^\dagger$ with $R$ real symmetric after every collision. Every bond-energy contribution $t_{ij}(C_{ij}+C_{ji})$ vanishes, and therefore
\begin{equation}
E_B=\epsilon_0N_B,\qquad
\langle H_{\mathrm{int}}\rangle=0,\qquad
W_{\mathrm{sw}}^{(\mathrm{ch})}=0.
\label{eq:resonant-identities}
\end{equation}
If $\omega_L=\epsilon_0$, the same form is preserved during discharge, so $\langle H_{BL}\rangle_t=0$ and $W_{\mathrm{sw}}^{(L)}=0$. These identities hold in both phases for any $N$, $n$, $g$, and $\tau$; they survive hopping disorder, dephasing, and uniform fermionic-mode loss, but are broken by on-site disorder or $\Delta_A\neq0$. Thus, for resonant charging, $\W_{\mathrm{net}}=\W_L$ and $\eta_{\mathrm{cyc}}=[2p_L-1]_+/N_B$.

A common shift by $\delta\epsilon$ of $\epsilon_0$, $\omega_A$, and $\omega_L$ multiplies each propagator by the commuting factor $e^{-i\delta\epsilon\widehat N_{\mathrm{tot}}t}$ and leaves $C_B$ unchanged at fixed detunings. Hence $\nu_{\max}$, $p_L$, the activation criterion, and $\W^{(2)}$ are independent of this shift. Global ergotropy is not, because unrestricted unitaries can reorder populations across excitation-number sectors.

\subsection{Numerical construction}

Small systems are propagated in the complete $2^{2N}$-dimensional battery Hilbert space. The memoryless boundary problem also admits an exact scalable free-fermion representation. The Jordan--Wigner transformation maps Eq.~\eqref{eq:HB} to a number-conserving quadratic Hamiltonian. Writing $u_1$ for the corresponding one-particle propagator when the charger addresses $A_1$ and $C_A=(1)$ for the excited-charger correlation matrix, one collision updates $C_B$ as
\begin{equation}
C'_B=[u_1(C_A\oplus C_B)u_1^\dagger]_B,
\label{eq:covariance}
\end{equation}
where $[\cdot]_B$ denotes the battery principal block. Vacuum input, occupied single-mode ancillas, number-conserving quadratic evolution, and partial traces preserve number-diagonal Gaussianity. If $r=\operatorname{rank}C_B$, the nonzero many-body state eigenvalues are $\lambda_{\bm s}=\prod_{q=1}^{r}\nu_q^{s_q}(1-\nu_q)^{1-s_q}$ with $s_q\in\{0,1\}$ \cite{Peschel2003}. Starting from vacuum, six fresh chargers imply $r\leq6$, so only the 64 lowest many-body energy subset sums are needed to evaluate Eq.~\eqref{eq:ergotropy}, even at $N=100$. The extraction step uses the same update, with the load as an additional mode adjacent to $A_1$. For noisy extraction, we propagate $C$ with a symmetric split-step scheme; dephasing and uniform loss act on $C$ in closed form \cite{Supplemental}.

\section{Results}

\subsection{Spectral matching creates an edge-orbital inversion}

At $N=100$ and resonance ($\Delta_A=0$), six collisions give $\nu_{\max}=0.869$ in the topological phase. The dominant natural orbital overlaps the finite-chain left-edge reference with weight $0.999$. In the trivial phase no one-particle state lies at $\epsilon_0$: the charger couples off-resonantly to the bulk bands, the excitation spreads over several dispersive orbitals, and $\nu_{\max}=0.418<1/2$.

Writing $r=v/w$ and $\mathcal N_N=[(1-r^2)/(1-r^{2N})]^{1/2}$, the charger couples to the left-edge mode with $g_{A,\mathrm{eff}}=g\mathcal N_N$. A resonant collision with a fresh excited charger updates the edge occupation as $\nu\mapsto\nu\cos^2\phi+\sin^2\phi$, where $\phi=g_{A,\mathrm{eff}}\tau$. Thus
\begin{equation}
\nu_{\mathrm{edge}}^{(n)}\simeq
1-\cos^{2n}\!\left(g\mathcal N_N\tau\right).
\label{eq:edge-filling}
\end{equation}
For $g=0.3J$, $\tau=2J^{-1}$, $r=1/3$, and $n=6$, Eq.~\eqref{eq:edge-filling} gives $0.869$. Inversion begins at $n\geq\lceil\ln2/|\ln\cos^2\phi|\rceil=3$: the load is passive for $n=1,2$ and reaches $\W_L\simeq0.67J$ at $n=3$. The reduction requires $g_{A,\mathrm{eff}}\ll|w-v|$; here these scales are $0.28J$ and $J$, respectively.

With $v=J(1-\delta)$ and $w=J(1+\delta)$, we scan $\delta$ across the SSH transition at $N=100$ with all controls fixed. Edge character develops first: the overlap of the dominant natural orbital with the edge reference orbital defined in the Supplemental Material crosses one half at $\delta\simeq0.028$. The natural occupation inverts at $\delta\simeq0.079$, and the prescribed boundary pulse activates the load at $\delta\simeq0.175$. The later load-activation crossover reflects the additional spatial and spectral overlap required by the fixed boundary pulse. The full sweep and the finite-size definition of the edge reference are given in Supplemental Fig.~S5 \cite{Supplemental}.

\begin{figure*}[!t]
  \centering
  \includegraphics{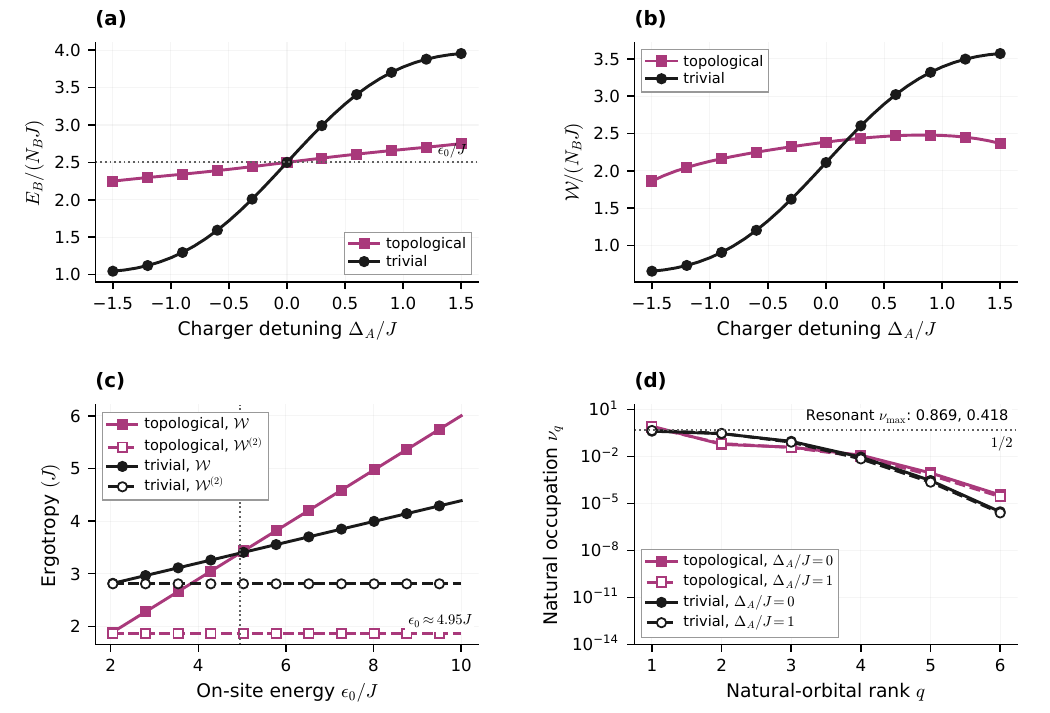}
  \caption{Spectral filtering and passive-state diagnostics after six memoryless collisions. Squares denote the topological phase and circles the trivial phase. (a) Stored energy and (b) global ergotropy per battery excitation versus charger detuning for $N=100$; at resonance, Eq.~\eqref{eq:resonant-identities} gives $E_B/N_B=\epsilon_0$ in both phases. (c) Global ergotropy $\W$ (solid) and number-conserving quadratic ergotropy $\W^{(2)}$ (dashed) versus a common shift of $\epsilon_0$ and $\omega_A$ at fixed $\Delta_A=J$; the vertical line marks the $N=100$ crossing of the global-ergotropy curves at $\epsilon_0\approx4.95J$, while $\W^{(2)}$ is invariant. (d) Six nonzero natural occupations at resonance and baseline: circles are trivial, squares are topological, solid lines use $\Delta_A=0$, and dashed lines use $\Delta_A=J$. At resonance, $\nu_{\max}=0.869$ (topological) and $0.418$ (trivial). The horizontal line is the inversion threshold $1/2$.}
  \label{fig:spectral}
\end{figure*}

At the baseline detuning, the trivial chain stores more energy and more global ergotropy per excitation than the topological chain [Fig.~\ref{fig:spectral}(a,b)]. Near resonance, the ranking of global ergotropy reverses. In both settings, only the topological load becomes active under the common pulse. Stored energy and global ergotropy therefore do not predict port-accessible work. This ranking depends on $\epsilon_0$ (Sec.~\ref{sec:work-measures}): the global-ergotropy curves cross at $\epsilon_0\approx4.95J$ [Fig.~\ref{fig:spectral}(c)], whereas $\W^{(2)}$ and the normalized load work $\W_L/\omega_L$ do not change.

\subsection{A boundary pulse nearly saturates the accessibility bound}

\begin{figure*}[!t]
  \centering
  \includegraphics{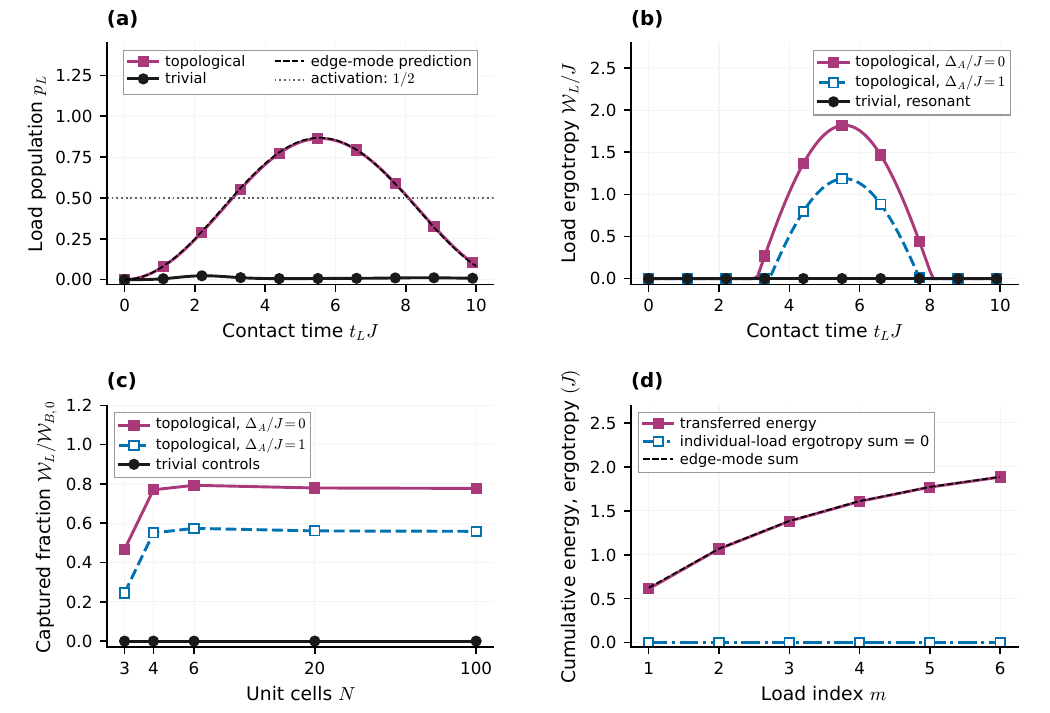}
  \caption{Local discharge after six fresh-charger collisions with $g_L=0.3J$ and $\omega_L=\epsilon_0$. Squares denote the topological phase and circles the trivial phase; in (b,c), solid lines with filled squares and dashed lines with open squares denote topological charging at $\Delta_A=0$ and $\Delta_A=J$, respectively. Thin black dashed curves are analytical predictions. (a) Load population versus contact time for $N=100$ after resonant charging; the dotted line is the active-load threshold and the dashed curve is Eq.~\eqref{eq:edge-load}. (b) For $N=100$, topological load ergotropy after resonant ($\Delta_A=0$) and baseline ($\Delta_A=J$) charging; the resonant trivial control remains zero. (c) Topological captured fraction $\W_L/\W_{B,0}$, with $\W_{B,0}$ the battery ergotropy before discharge, for $N=3,4,6,20,100$ on a logarithmic $N$ axis. At each $N$ and detuning, the topological-optimal pulse is applied unchanged to the trivial phase; all matched trivial loads are passive. (d) For $N=100$, six successive $t_L=2J^{-1}$ loads after resonant topological charging: solid with filled squares, cumulative transferred energy; dash-dotted with open squares, summed single-load ergotropy $\sum_{i=1}^{m}\W_{L,i}=0$; black dashed, Eq.~\eqref{eq:load-series}. Each short contact leaves its load below $p_{L,i}=1/2$, so the individual loads are passive despite the nonzero transferred energy. This sum excludes work accessible by joint operations on the loads; the edge-model collective load occupation reaches approximately $0.76$ and can be concentrated by a multi-load quadratic rotation.}
  \label{fig:load}
\end{figure*}

The load protocol uses $g_L=0.3J$, $\omega_L=\epsilon_0$, and one contact time selected by maximizing the topological load work. Both phases are evaluated with that same pulse. Figure~\ref{fig:load} shows the load trajectory, delivered work, the captured fraction $\W_L/\W_{B,0}$ of the battery ergotropy $\W_{B,0}$ before discharge across system sizes, and the cumulative response to successive short contacts.

For resonant topological charging at $N=100$, the optimized $t_L=5.6J^{-1}$ pulse gives $p_L=0.864$ and reaches $99.4\%$ of the population ceiling and $98.6\%$ of the corresponding work ceiling. The matched trivial load remains passive, as implied by $\nu_{\max}<1/2$. With $\nu_{\mathrm{edge}}$ the charged edge-orbital occupation, the isolated-edge reduction gives
\begin{equation}
\begin{aligned}
p_L(t)&\simeq\nu_{\mathrm{edge}}\sin^2(g_{L,\mathrm{eff}}t),\\
g_{L,\mathrm{eff}}&=g_L\mathcal N_N,\qquad
t_{\mathrm{swap}}=\frac{\pi}{2g_{L,\mathrm{eff}}}.
\end{aligned}
\label{eq:edge-load}
\end{equation}
For $r=1/3$, $N=100$, and $g_L=0.3J$, this predicts $t_{\mathrm{swap}}\simeq5.55J^{-1}$, consistent with the grid optimum $t_L=5.6J^{-1}$. Since $\mathcal N_N$ changes by less than $10^{-3}$ between $N=3$ and $N=100$, the same $t_L$ is near-optimal for all reported sizes. The residual population deficit is consistent with leakage into bulk states and has the estimated perturbative scale
\begin{equation}
1-\frac{p_L}{\nu_{\mathrm{edge}}}
\sim\frac{g_L^2r^2}{(w-v)^2}.
\label{eq:leakage}
\end{equation}
This estimate gives about $1\%$ for the reported controls, compared with the observed $0.6\%$ deficit; weaker load coupling or stronger dimerization brings the pulse closer to the bound. For $N\geq20$, the topological load captures about $0.78$ of the pre-discharge battery ergotropy after resonant charging and $0.56$ after baseline charging [Fig.~\ref{fig:load}(c)]. Resonant charging also yields the larger load work, $1.82J$ versus $1.19J$ [Fig.~\ref{fig:load}(b)]. For successive fresh loads of duration $t_s$, with $\theta_s=g_{L,\mathrm{eff}}t_s$, the same model gives $p_{L,m}\simeq\nu_{\mathrm{edge}}\sin^2\theta_s\cos^{2(m-1)}\theta_s$ and the cumulative transferred energy
\begin{equation}
E_L^{(k)}\simeq\omega_L\nu_{\mathrm{edge}}
\left(1-\cos^{2k}\theta_s\right).
\label{eq:load-series}
\end{equation}
For the short contacts $t_s=2J^{-1}$ shown in Fig.~\ref{fig:load}(d), each load remains passive, $p_{L,m}<1/2$. The transferred population, however, occupies a collective load mode with occupation $\nu_{\mathrm{edge}}(1-\cos^{2k}\theta_s)\simeq0.76$ for the six contacts shown. The work thus resides in a collective load mode. A quadratic rotation among the degenerate loads costs no energy and concentrates this occupation into one load. That load then holds $p_L\simeq0.76$ and yields $\W_L=\omega_L(2p_L-1)$.

\subsection{Delivered work and cycle power favor different charging durations}

\begin{figure*}[!t]
  \centering
  \begin{minipage}[t]{3.375in}
  \centering
  \includegraphics{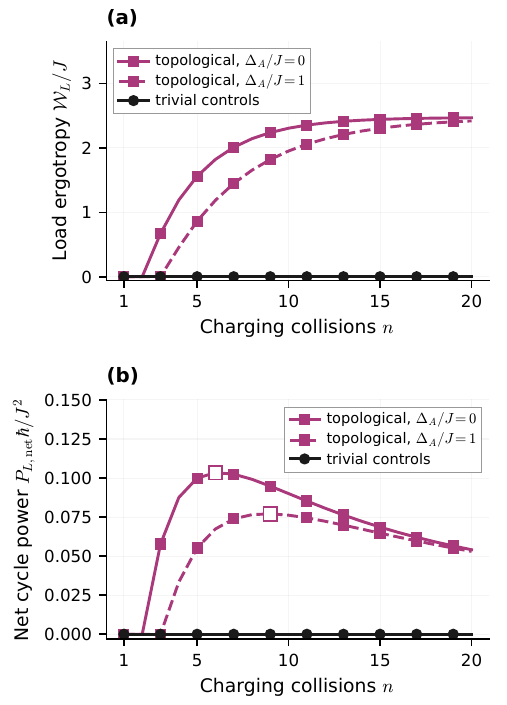}
  \caption{Cycle-level useful output for $N=20$ with $g_L=0.3J$ and $\omega_L=\epsilon_0$. Squares denote the topological phase and circles the trivial phase; solid and dashed lines denote $\Delta_A=0$ and $\Delta_A=J$, respectively. (a) Delivered load ergotropy and (b) zero-overhead net cycle power versus the number of charging collisions for resonant ($\Delta_A=0$) and baseline ($\Delta_A=J$) charging. At each $n$ and detuning, $t_L$ is optimized on a common grid to maximize the topological load ergotropy (or its population before activation) and applied unchanged to the trivial phase; since $t_{\mathrm{swap}}$ does not depend on $n$, the optimum stays near $5.6J^{-1}$. Larger open squares mark the power maxima; every trivial load remains passive.}
  \label{fig:cycle-performance}
  \end{minipage}
  \hfill
  \begin{minipage}[t]{3.375in}
  \centering
  \includegraphics{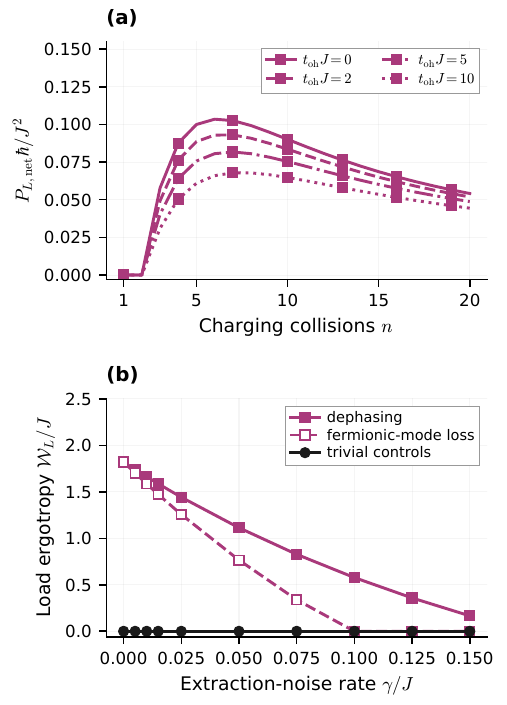}
  \caption{Cycle-power and extraction-noise controls for resonantly charged $N=20$ batteries. Squares denote topological results and circles the merged trivial controls. (a) Topological net cycle power for $t_{\mathrm{oh}}=0$, $2J^{-1}$, $5J^{-1}$, and $10J^{-1}$ (solid, dashed, dash-dotted, and dotted, respectively); switching costs vanish exactly at resonance by Eq.~\eqref{eq:resonant-identities}. At each collision count the load pulse optimized for delivered work in the topological phase is applied unchanged to the trivial phase. (b) Load ergotropy after six resonant collisions when independent dephasing (solid, filled squares) or fermionic-mode loss (dashed, open squares) acts on every battery mode and the load throughout the fixed $(g_L,t_L)=(0.3J,5.6J^{-1})$ extraction. Matched trivial loads remain passive. The loss curve obeys the exact uniform-loss law $p_L(\gamma)=e^{-\gamma t_L}p_L(0)$.}
  \label{fig:cycle-robustness}
  \end{minipage}
\end{figure*}

At $N=20$, we scan $n=1,\ldots,20$ charging collisions, optimizing $t_L$ for the topological phase at each $n$ on a common grid and applying the selected pulse unchanged to the trivial phase. The power maximum reported below is taken over collision number using these work-optimized load pulses. Joint optimization of collision number and load duration for power is outside the present protocol. Figure~\ref{fig:cycle-performance} shows the delivered load ergotropy and the zero-overhead net cycle power versus the number of collisions. The delivered work increases monotonically over the scanned range, whereas the net power passes through a maximum: resonant charging maximizes net load power after six collisions at $P_{L,\mathrm{net}}=0.103J^2/\hbar$, while baseline charging peaks after nine collisions at $0.077J^2/\hbar$. The six-collision resonant resource yield is $\eta_{\mathrm{cyc}}=0.741$. With $\kappa=p_L/\nu_{\mathrm{edge}}\simeq0.994$, the edge reduction gives
\begin{equation}
P_{L,\mathrm{net}}(n)\simeq
\frac{\omega_L[2\kappa\nu_{\mathrm{edge}}^{(n)}-1]_+}
{n\tau+t_L+t_{\mathrm{oh}}}.
\label{eq:power-approx}
\end{equation}
The load work saturates as $1-\cos^{2n}\phi$, whereas the cycle time grows linearly in $n$, so their ratio has a finite maximum. Prescribed overhead times $t_{\mathrm{oh}}=2J^{-1}$, $5J^{-1}$, and $10J^{-1}$ shift the optimum to larger $n$ and flatten the maximum [Fig.~\ref{fig:cycle-robustness}(a)]. The full data give $n^*=6$ for $t_{\mathrm{oh}}=0$ and $n^*=7$ for $t_{\mathrm{oh}}=2$, $5$, and $10J^{-1}$; at $2J^{-1}$ and $10J^{-1}$ the two best grid points differ by only $0.14\%$ and $0.11\%$, respectively.

\subsection{Noise channel and symmetry class determine robustness}

During extraction, we apply local dephasing or fermionic-mode loss to every battery mode and to the load. Dephasing conserves total excitation but damps the coherences that drive the swap. The load work therefore decreases gradually and remains $0.166J$ at $\gamma=0.15J$. Uniform fermionic-mode loss commutes with number-conserving quadratic dynamics and gives
\begin{equation}
C(t)=e^{-\gamma t}u(t)C(0)u^\dagger(t),\qquad
p_L(\gamma)=e^{-\gamma t_L}p_L(0).
\label{eq:loss-law}
\end{equation}
The load remains active for $\gamma<\gamma^*=\ln[2p_L(0)]/t_L\simeq0.098J$ [Fig.~\ref{fig:cycle-robustness}(b)].

\begin{figure*}[!t]
  \centering
  \includegraphics{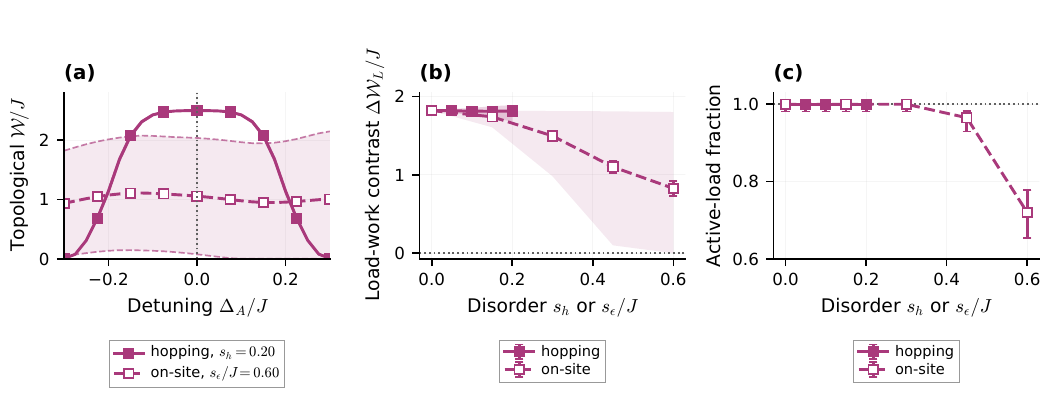}
  \caption{Symmetry-resolved disorder for $N=20$. (a) Long-contact edge spectroscopy after six collisions with $(g,\tau)=(0.05J,20J^{-1})$. Filled squares with solid lines denote hopping disorder $s_h=0.20$, and open squares with dashed lines denote on-site disorder $s_\epsilon=0.60J$; curves and bands show the mean topological global ergotropy and one sample standard deviation over 200 realizations. The hopping-disorder band is narrow on this scale. Individual edge-centroid shifts are reported in the Supplemental Material \cite{Supplemental}. Panels (b,c) use the common resonant load pulse $(g_L,t_L)=(0.3J,5.6J^{-1})$ and 200 paired realizations at each disorder strength. (b) Mean paired load-work contrast with two-sided Student-$t$ 95\% intervals; shading spans the empirical fifth--95th percentiles. (c) Fraction of active topological loads with Wilson 95\% intervals \cite{Wilson1927}. Every matched trivial load remains passive.}
  \label{fig:disorder}
\end{figure*}

We classify static disorder by chiral symmetry. Every realization retains the clean-system charger frequency and load pulse. We sample $v_j=v(1+s_hx_j)$, $w_j=w(1+s_hy_j)$, and $\epsilon_\ell=\epsilon_0+s_\epsilon z_\ell$, with independent $x_j,y_j,z_\ell$ uniform on $[-1,1]$. Thus $s_h$ is dimensionless and $s_\epsilon$ is expressed in $J$. Hopping disorder preserves the sublattice structure of $H_B-\epsilon_0I$; on-site disorder breaks it. The long-contact spectroscopy calculation uses independent phase-resolved ensembles at $N=20$, six collisions, $(g,\tau)=(0.05J,20J^{-1})$, 25 detunings on $[-0.3J,0.3J]$, and 200 realizations per phase and disorder class. The fixed-pulse discharge calculation separately pairs the same normalized draw between phases in each of 200 realizations. Weak, long collisions resolve the edge energy with a narrow spectral window. Figure~\ref{fig:disorder} summarizes the ensemble spectra and fixed-control load statistics, while realization-resolved centroid shifts are given in the Supplemental Material \cite{Supplemental}.

For hopping disorder that leaves the chain in the topological regime (here $|v_j/w_j|\leq0.5$ on every bond), chiral symmetry keeps the semi-infinite edge mode at $\epsilon_0$ with a normalizable profile, $\alpha_{j+1}=-(v_j/w_j)\alpha_j$. Hopping disorder changes the normalization and hence the effective edge--load coupling, while the clean $t_L$ lies near the swap maximum, suppressing the first-order sensitivity of $\sin^2(g_{L,\mathrm{eff}}t_L)$ to coupling variations. At $t_L=t_{\mathrm{swap}}$, this factor changes only at second order. On-site disorder instead shifts the edge energy at first order, $\delta E_{\mathrm{edge}}\simeq s_\epsilon\sum_j|\alpha_j|^2z_{A_j}$. For the clean semi-infinite profile its standard deviation is
\begin{equation}
\sigma_E=s_\epsilon\left[\frac{1-r^2}{3(1+r^2)}\right]^{1/2}
\simeq0.52s_\epsilon.
\label{eq:disorder-width}
\end{equation}
Thus $\sigma_E\simeq0.31J$ at $s_\epsilon=0.6J$, comparable to the standard charging linewidth $2g_{A,\mathrm{eff}}\simeq0.57J$. This first-order detuning is consistent with the passive-load tail.

For 200 matched realizations at hopping-disorder strength $s_h=0.20$, the mean topological-minus-trivial load-work contrast is $1.815J$ with a two-sided Student-$t$ 95\% confidence interval $[1.806,1.824]J$, and every topological load is active. At on-site half-width $s_\epsilon=0.60J$, the active fraction falls to $0.72$ and the fifth contrast percentile reaches zero. The positive mean contrast coexists with a finite failure probability under strong on-site disorder.

\section{Discussion}

The results follow a three-step mechanism. The edge mode localizes the charge at the port, resonant collisions invert its occupation, and Eq.~\eqref{eq:bound} fixes how much of it a local quadratic protocol can extract.

Earlier SSH-battery studies evaluated charging and global ergotropy \cite{Zhao2022}, and topological waveguides have been used to mediate charger--battery transfer \cite{Lu2025}. Here the SSH chain is itself the battery, and the figure of merit is the work delivered through one port.

Equation~\eqref{eq:bound} applies to number-conserving quadratic mode rotations and directly bounds the single-load protocol studied here. Collective many-copy extraction and correlation-assisted transport can access resources beyond this single-load setting \cite{Andolina2019,Simon2025}. Interacting boundary controls also fall outside the quadratic class. The reported cycle power and resource yield use the prescribed overhead and ideal load-work extraction conventions. The open-system calculations describe phenomenological dephasing and fermionic-mode loss. Local dephasing admits a closed evolution equation for the one-body correlation matrix, which we integrate using a timestep-converged symmetric split-step scheme. Local qubit loss coincides with fermionic-mode loss in the zero- and one-excitation sectors; deviations are controlled by the multi-excitation weight of the state \cite{Supplemental}. Dense-system, control-noise, and additional open-system checks are given in the Supplemental Material \cite{Supplemental}.

\section{Conclusion}

We asked when energy stored in a structured quantum battery becomes work at a single local port. We studied a hard-core SSH chain charged at its boundary by repeated collisions with excited two-level ancillas. An initially empty two-level load then discharged the chain through the same site. A free-fermion description gave a rigorous bound for all number-conserving quadratic extraction protocols. The load population cannot exceed the largest natural occupation of the battery, and a quadratic swap attains it, so the load gains work exactly when this occupation exceeds one half. The population bound holds for any battery state and any chain length; the work bound assumes a number-diagonal battery state.

Two analytic results explain how the topological chain reaches this bound. Resonant collisions fill the edge mode geometrically, $\nu_{\mathrm{edge}}^{(n)}\simeq1-\cos^{2n}(g\mathcal N_N\tau)$, so inversion requires only a finite number of collisions set by the charger--edge coupling. A single boundary pulse then swaps this occupation into the load. Its small deficit from the bound is consistent with leakage into bulk states, with the estimated perturbative scale $g_L^2r^2/(w-v)^2$. For the representative parameters, the pulse delivers $98.6\%$ of the maximal load work. The trivial chain has no midgap state. Over the charging horizons studied, its excitation stays spread over bulk orbitals and its load remains passive, although it stores more energy at the baseline detuning. A sweep through the SSH transition shows that edge localization, occupation inversion, and load activation set in successively. Within the work-optimized pulse protocol, load work approaches saturation as the number of collisions increases, whereas the cycle time grows linearly, producing a finite maximum of the cycle power. Hopping disorder preserves chiral symmetry and the midgap edge channel, and the work contrast persists over the tested range. On-site disorder shifts the edge energy at first order and leaves a finite fraction of loads passive.

These results distinguish three quantities: stored energy, global ergotropy, and work available at a port. Topology supplies a localized, symmetry-protected channel; resonance and the available control class decide whether this channel delivers work. Because the population bound depends only on the one-body correlation matrix, it applies to any battery with number-conserving quadratic dynamics and a local port. Interacting controls and extraction into multiple loads extend the single-load quadratic setting. Quantifying the work available under these broader controls is a natural next step. Higher-dimensional batteries with corner or hinge modes offer a further extension. Programmable superconducting qubit chains offer a near-term platform to test these predictions and to explore local work extraction from topological matter.

\begin{acknowledgments}
The authors used generative AI tools (OpenAI ChatGPT and Anthropic Claude) for presentation and language editing. The authors reviewed all edits and take full responsibility for the content.
\end{acknowledgments}

\section*{Data Availability}
The numerical data and source code supporting the findings of this study are available from the corresponding author upon reasonable request.

\clearpage
\bibliography{references}

\begin{thebibliography}{30}%
\makeatletter
\providecommand \@ifxundefined [1]{%
 \@ifx{#1\undefined}
}%
\providecommand \@ifnum [1]{%
 \ifnum #1\expandafter \@firstoftwo
 \else \expandafter \@secondoftwo
 \fi
}%
\providecommand \@ifx [1]{%
 \ifx #1\expandafter \@firstoftwo
 \else \expandafter \@secondoftwo
 \fi
}%
\providecommand \natexlab [1]{#1}%
\providecommand \enquote  [1]{``#1''}%
\providecommand \bibnamefont  [1]{#1}%
\providecommand \bibfnamefont [1]{#1}%
\providecommand \citenamefont [1]{#1}%
\providecommand \href@noop [0]{\@secondoftwo}%
\providecommand \href [0]{\begingroup \@sanitize@url \@href}%
\providecommand \@href[1]{\@@startlink{#1}\@@href}%
\providecommand \@@href[1]{\endgroup#1\@@endlink}%
\providecommand \@sanitize@url [0]{\catcode `\\12\catcode `\$12\catcode
  `\&12\catcode `\#12\catcode `\^12\catcode `\_12\catcode `\%12\relax}%
\providecommand \@@startlink[1]{}%
\providecommand \@@endlink[0]{}%
\providecommand \url  [0]{\begingroup\@sanitize@url \@url }%
\providecommand \@url [1]{\endgroup\@href {#1}{\urlprefix }}%
\providecommand \urlprefix  [0]{URL }%
\providecommand \Eprint [0]{\href }%
\providecommand \doibase [0]{https://doi.org/}%
\providecommand \selectlanguage [0]{\@gobble}%
\providecommand \bibinfo  [0]{\@secondoftwo}%
\providecommand \bibfield  [0]{\@secondoftwo}%
\providecommand \translation [1]{[#1]}%
\providecommand \BibitemOpen [0]{}%
\providecommand \bibitemStop [0]{}%
\providecommand \bibitemNoStop [0]{.\EOS\space}%
\providecommand \EOS [0]{\spacefactor3000\relax}%
\providecommand \BibitemShut  [1]{\csname bibitem#1\endcsname}%
\let\auto@bib@innerbib\@empty
\bibitem [{\citenamefont {Allahverdyan}\ \emph {et~al.}(2004)\citenamefont
  {Allahverdyan}, \citenamefont {Balian},\ and\ \citenamefont {{Th. M.
  Nieuwenhuizen}}}]{Allahverdyan2004}%
  \BibitemOpen
  \bibfield  {author} {\bibinfo {author} {\bibfnamefont {A.~E.}\ \bibnamefont
  {Allahverdyan}}, \bibinfo {author} {\bibfnamefont {R.}~\bibnamefont
  {Balian}},\ and\ \bibinfo {author} {\bibnamefont {{Th. M. Nieuwenhuizen}}},\
  }\bibfield  {title} {\bibinfo {title} {Maximal work extraction from finite
  quantum systems},\ }\href {https://doi.org/10.1209/epl/i2004-10101-2}
  {\bibfield  {journal} {\bibinfo  {journal} {Europhys. Lett.}\ }\textbf
  {\bibinfo {volume} {67}},\ \bibinfo {pages} {565} (\bibinfo {year}
  {2004})}\BibitemShut {NoStop}%
\bibitem [{\citenamefont {Vinjanampathy}\ and\ \citenamefont
  {Anders}(2016)}]{Vinjanampathy2016}%
  \BibitemOpen
  \bibfield  {author} {\bibinfo {author} {\bibfnamefont {S.}~\bibnamefont
  {Vinjanampathy}}\ and\ \bibinfo {author} {\bibfnamefont {J.}~\bibnamefont
  {Anders}},\ }\bibfield  {title} {\bibinfo {title} {Quantum thermodynamics},\
  }\href {https://doi.org/10.1080/00107514.2016.1201896} {\bibfield  {journal}
  {\bibinfo  {journal} {Contemp. Phys.}\ }\textbf {\bibinfo {volume} {57}},\
  \bibinfo {pages} {545} (\bibinfo {year} {2016})}\BibitemShut {NoStop}%
\bibitem [{\citenamefont {Campaioli}\ \emph {et~al.}(2024)\citenamefont
  {Campaioli}, \citenamefont {Gherardini}, \citenamefont {Quach}, \citenamefont
  {Polini},\ and\ \citenamefont {Andolina}}]{Campaioli2024}%
  \BibitemOpen
  \bibfield  {author} {\bibinfo {author} {\bibfnamefont {F.}~\bibnamefont
  {Campaioli}}, \bibinfo {author} {\bibfnamefont {S.}~\bibnamefont
  {Gherardini}}, \bibinfo {author} {\bibfnamefont {J.~Q.}\ \bibnamefont
  {Quach}}, \bibinfo {author} {\bibfnamefont {M.}~\bibnamefont {Polini}},\ and\
  \bibinfo {author} {\bibfnamefont {G.~M.}\ \bibnamefont {Andolina}},\
  }\bibfield  {title} {\bibinfo {title} {Colloquium: Quantum batteries},\
  }\href {https://doi.org/10.1103/RevModPhys.96.031001} {\bibfield  {journal}
  {\bibinfo  {journal} {Rev. Mod. Phys.}\ }\textbf {\bibinfo {volume} {96}},\
  \bibinfo {pages} {031001} (\bibinfo {year} {2024})}\BibitemShut {NoStop}%
\bibitem [{\citenamefont {Francica}\ \emph {et~al.}(2020)\citenamefont
  {Francica}, \citenamefont {Binder}, \citenamefont {Guarnieri}, \citenamefont
  {Mitchison}, \citenamefont {Goold},\ and\ \citenamefont
  {Plastina}}]{Francica2020}%
  \BibitemOpen
  \bibfield  {author} {\bibinfo {author} {\bibfnamefont {G.}~\bibnamefont
  {Francica}}, \bibinfo {author} {\bibfnamefont {F.~C.}\ \bibnamefont
  {Binder}}, \bibinfo {author} {\bibfnamefont {G.}~\bibnamefont {Guarnieri}},
  \bibinfo {author} {\bibfnamefont {M.~T.}\ \bibnamefont {Mitchison}}, \bibinfo
  {author} {\bibfnamefont {J.}~\bibnamefont {Goold}},\ and\ \bibinfo {author}
  {\bibfnamefont {F.}~\bibnamefont {Plastina}},\ }\bibfield  {title} {\bibinfo
  {title} {Quantum coherence and ergotropy},\ }\href
  {https://doi.org/10.1103/PhysRevLett.125.180603} {\bibfield  {journal}
  {\bibinfo  {journal} {Phys. Rev. Lett.}\ }\textbf {\bibinfo {volume} {125}},\
  \bibinfo {pages} {180603} (\bibinfo {year} {2020})}\BibitemShut {NoStop}%
\bibitem [{\citenamefont {Simon}\ \emph {et~al.}(2025)\citenamefont {Simon},
  \citenamefont {Anders},\ and\ \citenamefont {Hovhannisyan}}]{Simon2025}%
  \BibitemOpen
  \bibfield  {author} {\bibinfo {author} {\bibfnamefont {R.~P.~A.}\
  \bibnamefont {Simon}}, \bibinfo {author} {\bibfnamefont {J.}~\bibnamefont
  {Anders}},\ and\ \bibinfo {author} {\bibfnamefont {K.~V.}\ \bibnamefont
  {Hovhannisyan}},\ }\bibfield  {title} {\bibinfo {title} {Correlations enable
  lossless ergotropy transport},\ }\href
  {https://doi.org/10.1103/PhysRevLett.134.010408} {\bibfield  {journal}
  {\bibinfo  {journal} {Phys. Rev. Lett.}\ }\textbf {\bibinfo {volume} {134}},\
  \bibinfo {pages} {010408} (\bibinfo {year} {2025})}\BibitemShut {NoStop}%
\bibitem [{\citenamefont {Ferraro}\ \emph {et~al.}(2026)\citenamefont
  {Ferraro}, \citenamefont {Cavaliere}, \citenamefont {Genoni}, \citenamefont
  {Benenti},\ and\ \citenamefont {Sassetti}}]{Ferraro2026Perspective}%
  \BibitemOpen
  \bibfield  {author} {\bibinfo {author} {\bibfnamefont {D.}~\bibnamefont
  {Ferraro}}, \bibinfo {author} {\bibfnamefont {F.}~\bibnamefont {Cavaliere}},
  \bibinfo {author} {\bibfnamefont {M.~G.}\ \bibnamefont {Genoni}}, \bibinfo
  {author} {\bibfnamefont {G.}~\bibnamefont {Benenti}},\ and\ \bibinfo {author}
  {\bibfnamefont {M.}~\bibnamefont {Sassetti}},\ }\bibfield  {title} {\bibinfo
  {title} {Opportunities and challenges of quantum batteries},\ }\href
  {https://doi.org/10.1038/s42254-025-00906-5} {\bibfield  {journal} {\bibinfo
  {journal} {Nat. Rev. Phys.}\ }\textbf {\bibinfo {volume} {8}},\ \bibinfo
  {pages} {115} (\bibinfo {year} {2026})}\BibitemShut {NoStop}%
\bibitem [{\citenamefont {Hu}\ \emph {et~al.}(2026)\citenamefont {Hu},
  \citenamefont {Liu}, \citenamefont {Zhao}, \citenamefont {Zhong},
  \citenamefont {Zhou}, \citenamefont {Liu}, \citenamefont {Yuan},
  \citenamefont {Lin}, \citenamefont {Xu}, \citenamefont {Hu}, \citenamefont
  {Xie}, \citenamefont {Liu}, \citenamefont {Zhou}, \citenamefont {Ri},
  \citenamefont {Zhang}, \citenamefont {Deng}, \citenamefont {Saguia},
  \citenamefont {Linpeng}, \citenamefont {Sarandy}, \citenamefont {Liu},
  \citenamefont {Santos}, \citenamefont {Tan},\ and\ \citenamefont
  {Yu}}]{Hu2026Charging}%
  \BibitemOpen
  \bibfield  {author} {\bibinfo {author} {\bibfnamefont {C.-K.}\ \bibnamefont
  {Hu}}, \bibinfo {author} {\bibfnamefont {C.}~\bibnamefont {Liu}}, \bibinfo
  {author} {\bibfnamefont {J.}~\bibnamefont {Zhao}}, \bibinfo {author}
  {\bibfnamefont {L.}~\bibnamefont {Zhong}}, \bibinfo {author} {\bibfnamefont
  {Y.}~\bibnamefont {Zhou}}, \bibinfo {author} {\bibfnamefont {M.}~\bibnamefont
  {Liu}}, \bibinfo {author} {\bibfnamefont {H.}~\bibnamefont {Yuan}}, \bibinfo
  {author} {\bibfnamefont {Y.}~\bibnamefont {Lin}}, \bibinfo {author}
  {\bibfnamefont {Y.}~\bibnamefont {Xu}}, \bibinfo {author} {\bibfnamefont
  {G.}~\bibnamefont {Hu}}, \bibinfo {author} {\bibfnamefont {G.}~\bibnamefont
  {Xie}}, \bibinfo {author} {\bibfnamefont {Z.}~\bibnamefont {Liu}}, \bibinfo
  {author} {\bibfnamefont {R.}~\bibnamefont {Zhou}}, \bibinfo {author}
  {\bibfnamefont {Y.}~\bibnamefont {Ri}}, \bibinfo {author} {\bibfnamefont
  {W.}~\bibnamefont {Zhang}}, \bibinfo {author} {\bibfnamefont
  {R.}~\bibnamefont {Deng}}, \bibinfo {author} {\bibfnamefont {A.}~\bibnamefont
  {Saguia}}, \bibinfo {author} {\bibfnamefont {X.}~\bibnamefont {Linpeng}},
  \bibinfo {author} {\bibfnamefont {M.~S.}\ \bibnamefont {Sarandy}}, \bibinfo
  {author} {\bibfnamefont {S.}~\bibnamefont {Liu}}, \bibinfo {author}
  {\bibfnamefont {A.~C.}\ \bibnamefont {Santos}}, \bibinfo {author}
  {\bibfnamefont {D.}~\bibnamefont {Tan}},\ and\ \bibinfo {author}
  {\bibfnamefont {D.}~\bibnamefont {Yu}},\ }\bibfield  {title} {\bibinfo
  {title} {Quantum charging advantage in superconducting solid-state
  batteries},\ }\href {https://doi.org/10.1103/sp5l-c6m8} {\bibfield  {journal}
  {\bibinfo  {journal} {Phys. Rev. Lett.}\ }\textbf {\bibinfo {volume} {136}},\
  \bibinfo {pages} {060401} (\bibinfo {year} {2026})}\BibitemShut {NoStop}%
\bibitem [{\citenamefont {Yan}\ \emph {et~al.}(2026)\citenamefont {Yan},
  \citenamefont {Liu}, \citenamefont {Xiao}, \citenamefont {Lin}, \citenamefont
  {Song}, \citenamefont {Chen}, \citenamefont {Nori},\ and\ \citenamefont
  {Xia}}]{Yan2026Giant}%
  \BibitemOpen
  \bibfield  {author} {\bibinfo {author} {\bibfnamefont {K.-X.}\ \bibnamefont
  {Yan}}, \bibinfo {author} {\bibfnamefont {Y.}~\bibnamefont {Liu}}, \bibinfo
  {author} {\bibfnamefont {Y.}~\bibnamefont {Xiao}}, \bibinfo {author}
  {\bibfnamefont {J.-H.}\ \bibnamefont {Lin}}, \bibinfo {author} {\bibfnamefont
  {J.}~\bibnamefont {Song}}, \bibinfo {author} {\bibfnamefont {Y.-H.}\
  \bibnamefont {Chen}}, \bibinfo {author} {\bibfnamefont {F.}~\bibnamefont
  {Nori}},\ and\ \bibinfo {author} {\bibfnamefont {Y.}~\bibnamefont {Xia}},\
  }\bibfield  {title} {\bibinfo {title} {Giant-atom quantum batteries: Lossless
  energy transfer via interference engineering},\ }\href
  {https://doi.org/10.1103/43n6-rnj3} {\bibfield  {journal} {\bibinfo
  {journal} {Phys. Rev. Lett.}\ }\textbf {\bibinfo {volume} {136}},\ \bibinfo
  {pages} {180401} (\bibinfo {year} {2026})}\BibitemShut {NoStop}%
\bibitem [{\citenamefont {Su}\ \emph {et~al.}(1979)\citenamefont {Su},
  \citenamefont {Schrieffer},\ and\ \citenamefont {Heeger}}]{Su1979}%
  \BibitemOpen
  \bibfield  {author} {\bibinfo {author} {\bibfnamefont {W.~P.}\ \bibnamefont
  {Su}}, \bibinfo {author} {\bibfnamefont {J.~R.}\ \bibnamefont {Schrieffer}},\
  and\ \bibinfo {author} {\bibfnamefont {A.~J.}\ \bibnamefont {Heeger}},\
  }\bibfield  {title} {\bibinfo {title} {Solitons in polyacetylene},\ }\href
  {https://doi.org/10.1103/PhysRevLett.42.1698} {\bibfield  {journal} {\bibinfo
   {journal} {Phys. Rev. Lett.}\ }\textbf {\bibinfo {volume} {42}},\ \bibinfo
  {pages} {1698} (\bibinfo {year} {1979})}\BibitemShut {NoStop}%
\bibitem [{\citenamefont {Ryu}\ and\ \citenamefont {Hatsugai}(2002)}]{Ryu2002}%
  \BibitemOpen
  \bibfield  {author} {\bibinfo {author} {\bibfnamefont {S.}~\bibnamefont
  {Ryu}}\ and\ \bibinfo {author} {\bibfnamefont {Y.}~\bibnamefont {Hatsugai}},\
  }\bibfield  {title} {\bibinfo {title} {Topological origin of zero-energy edge
  states in particle-hole symmetric systems},\ }\href
  {https://doi.org/10.1103/PhysRevLett.89.077002} {\bibfield  {journal}
  {\bibinfo  {journal} {Phys. Rev. Lett.}\ }\textbf {\bibinfo {volume} {89}},\
  \bibinfo {pages} {077002} (\bibinfo {year} {2002})}\BibitemShut {NoStop}%
\bibitem [{\citenamefont {Asb\'oth}\ \emph {et~al.}(2016)\citenamefont
  {Asb\'oth}, \citenamefont {Oroszl\'any},\ and\ \citenamefont
  {P\'alyi}}]{Asboth2016}%
  \BibitemOpen
  \bibfield  {author} {\bibinfo {author} {\bibfnamefont {J.~K.}\ \bibnamefont
  {Asb\'oth}}, \bibinfo {author} {\bibfnamefont {L.}~\bibnamefont
  {Oroszl\'any}},\ and\ \bibinfo {author} {\bibfnamefont {A.}~\bibnamefont
  {P\'alyi}},\ }\href {https://doi.org/10.1007/978-3-319-25607-8} {\emph
  {\bibinfo {title} {A Short Course on Topological Insulators}}}\ (\bibinfo
  {publisher} {Springer},\ \bibinfo {address} {Cham},\ \bibinfo {year}
  {2016})\BibitemShut {NoStop}%
\bibitem [{\citenamefont {Zhao}\ \emph {et~al.}(2022)\citenamefont {Zhao},
  \citenamefont {Dou},\ and\ \citenamefont {Zhao}}]{Zhao2022}%
  \BibitemOpen
  \bibfield  {author} {\bibinfo {author} {\bibfnamefont {F.}~\bibnamefont
  {Zhao}}, \bibinfo {author} {\bibfnamefont {F.-Q.}\ \bibnamefont {Dou}},\ and\
  \bibinfo {author} {\bibfnamefont {Q.}~\bibnamefont {Zhao}},\ }\bibfield
  {title} {\bibinfo {title} {Charging performance of the
  {Su--Schrieffer--Heeger} quantum battery},\ }\href
  {https://doi.org/10.1103/PhysRevResearch.4.013172} {\bibfield  {journal}
  {\bibinfo  {journal} {Phys. Rev. Research}\ }\textbf {\bibinfo {volume}
  {4}},\ \bibinfo {pages} {013172} (\bibinfo {year} {2022})}\BibitemShut
  {NoStop}%
\bibitem [{\citenamefont {Lu}\ \emph {et~al.}(2025)\citenamefont {Lu},
  \citenamefont {Tian}, \citenamefont {L\"u},\ and\ \citenamefont
  {Shang}}]{Lu2025}%
  \BibitemOpen
  \bibfield  {author} {\bibinfo {author} {\bibfnamefont {Z.-G.}\ \bibnamefont
  {Lu}}, \bibinfo {author} {\bibfnamefont {G.}~\bibnamefont {Tian}}, \bibinfo
  {author} {\bibfnamefont {X.-Y.}\ \bibnamefont {L\"u}},\ and\ \bibinfo
  {author} {\bibfnamefont {C.}~\bibnamefont {Shang}},\ }\bibfield  {title}
  {\bibinfo {title} {Topological quantum batteries},\ }\href
  {https://doi.org/10.1103/PhysRevLett.134.180401} {\bibfield  {journal}
  {\bibinfo  {journal} {Phys. Rev. Lett.}\ }\textbf {\bibinfo {volume} {134}},\
  \bibinfo {pages} {180401} (\bibinfo {year} {2025})}\BibitemShut {NoStop}%
\bibitem [{\citenamefont {Kheiri}\ \emph {et~al.}(2026)\citenamefont {Kheiri},
  \citenamefont {Kheiri},\ and\ \citenamefont {Mahdavifar}}]{Kheiri2026}%
  \BibitemOpen
  \bibfield  {author} {\bibinfo {author} {\bibfnamefont {S.}~\bibnamefont
  {Kheiri}}, \bibinfo {author} {\bibfnamefont {S.}~\bibnamefont {Kheiri}},\
  and\ \bibinfo {author} {\bibfnamefont {S.}~\bibnamefont {Mahdavifar}},\
  }\bibfield  {title} {\bibinfo {title} {Dynamical phase transitions and
  criticality-enhanced charging in a topological quantum battery},\ }\href
  {https://doi.org/10.1103/42rw-l31q} {\bibfield  {journal} {\bibinfo
  {journal} {Phys. Rev. B}\ }\textbf {\bibinfo {volume} {113}},\ \bibinfo
  {pages} {205150} (\bibinfo {year} {2026})}\BibitemShut {NoStop}%
\bibitem [{\citenamefont {Yang}\ \emph {et~al.}(2026)\citenamefont {Yang},
  \citenamefont {Xiong}, \citenamefont {Liu}, \citenamefont {Yu}, \citenamefont
  {Liu}, \citenamefont {Mei},\ and\ \citenamefont {Shan}}]{Yang2026}%
  \BibitemOpen
  \bibfield  {author} {\bibinfo {author} {\bibfnamefont {J.}~\bibnamefont
  {Yang}}, \bibinfo {author} {\bibfnamefont {B.}~\bibnamefont {Xiong}},
  \bibinfo {author} {\bibfnamefont {J.}~\bibnamefont {Liu}}, \bibinfo {author}
  {\bibfnamefont {H.}~\bibnamefont {Yu}}, \bibinfo {author} {\bibfnamefont
  {D.}~\bibnamefont {Liu}}, \bibinfo {author} {\bibfnamefont {F.}~\bibnamefont
  {Mei}},\ and\ \bibinfo {author} {\bibfnamefont {C.}~\bibnamefont {Shan}},\
  }\href {https://doi.org/10.48550/arXiv.2609.03543} {\bibinfo {title}
  {Pulse-controlled topologically protected quantum batteries}} (\bibinfo
  {year} {2026}),\ \Eprint {https://arxiv.org/abs/2609.03543} {arXiv:2609.03543
  [quant-ph]} \BibitemShut {NoStop}%
\bibitem [{\citenamefont {Zhou}\ \emph {et~al.}(2026)\citenamefont {Zhou},
  \citenamefont {Xiao}, \citenamefont {Xu}, \citenamefont {Gao}, \citenamefont
  {Li}, \citenamefont {Zhou}, \citenamefont {Li},\ and\ \citenamefont
  {Shu}}]{Zhou2026}%
  \BibitemOpen
  \bibfield  {author} {\bibinfo {author} {\bibfnamefont {A.-L.}\ \bibnamefont
  {Zhou}}, \bibinfo {author} {\bibfnamefont {Y.-W.}\ \bibnamefont {Xiao}},
  \bibinfo {author} {\bibfnamefont {N.}~\bibnamefont {Xu}}, \bibinfo {author}
  {\bibfnamefont {L.-L.}\ \bibnamefont {Gao}}, \bibinfo {author} {\bibfnamefont
  {L.-J.}\ \bibnamefont {Li}}, \bibinfo {author} {\bibfnamefont
  {H.}~\bibnamefont {Zhou}}, \bibinfo {author} {\bibfnamefont {Z.-M.}\
  \bibnamefont {Li}},\ and\ \bibinfo {author} {\bibfnamefont {C.-C.}\
  \bibnamefont {Shu}},\ }\bibfield  {title} {\bibinfo {title} {Topological
  enhancement of a {$\mathcal{PT}$}-symmetric {Su--Schrieffer--Heeger} quantum
  battery},\ }\href {https://doi.org/10.1103/4klp-kw27} {\bibfield  {journal}
  {\bibinfo  {journal} {Phys. Rev. A}\ }\textbf {\bibinfo {volume} {113}},\
  \bibinfo {pages} {042213} (\bibinfo {year} {2026})}\BibitemShut {NoStop}%
\bibitem [{\citenamefont {Seah}\ \emph {et~al.}(2021)\citenamefont {Seah},
  \citenamefont {Perarnau-Llobet}, \citenamefont {Haack}, \citenamefont
  {Brunner},\ and\ \citenamefont {Nimmrichter}}]{Seah2021}%
  \BibitemOpen
  \bibfield  {author} {\bibinfo {author} {\bibfnamefont {S.}~\bibnamefont
  {Seah}}, \bibinfo {author} {\bibfnamefont {M.}~\bibnamefont
  {Perarnau-Llobet}}, \bibinfo {author} {\bibfnamefont {G.}~\bibnamefont
  {Haack}}, \bibinfo {author} {\bibfnamefont {N.}~\bibnamefont {Brunner}},\
  and\ \bibinfo {author} {\bibfnamefont {S.}~\bibnamefont {Nimmrichter}},\
  }\bibfield  {title} {\bibinfo {title} {Quantum speed-up in collisional
  battery charging},\ }\href {https://doi.org/10.1103/PhysRevLett.127.100601}
  {\bibfield  {journal} {\bibinfo  {journal} {Phys. Rev. Lett.}\ }\textbf
  {\bibinfo {volume} {127}},\ \bibinfo {pages} {100601} (\bibinfo {year}
  {2021})}\BibitemShut {NoStop}%
\bibitem [{\citenamefont {Salvia}\ \emph {et~al.}(2023)\citenamefont {Salvia},
  \citenamefont {Perarnau-Llobet}, \citenamefont {Haack}, \citenamefont
  {Brunner},\ and\ \citenamefont {Nimmrichter}}]{Salvia2023}%
  \BibitemOpen
  \bibfield  {author} {\bibinfo {author} {\bibfnamefont {R.}~\bibnamefont
  {Salvia}}, \bibinfo {author} {\bibfnamefont {M.}~\bibnamefont
  {Perarnau-Llobet}}, \bibinfo {author} {\bibfnamefont {G.}~\bibnamefont
  {Haack}}, \bibinfo {author} {\bibfnamefont {N.}~\bibnamefont {Brunner}},\
  and\ \bibinfo {author} {\bibfnamefont {S.}~\bibnamefont {Nimmrichter}},\
  }\bibfield  {title} {\bibinfo {title} {Quantum advantage in charging cavity
  and spin batteries by repeated interactions},\ }\href
  {https://doi.org/10.1103/PhysRevResearch.5.013155} {\bibfield  {journal}
  {\bibinfo  {journal} {Phys. Rev. Research}\ }\textbf {\bibinfo {volume}
  {5}},\ \bibinfo {pages} {013155} (\bibinfo {year} {2023})}\BibitemShut
  {NoStop}%
\bibitem [{\citenamefont {Catalano}\ \emph {et~al.}(2024)\citenamefont
  {Catalano}, \citenamefont {Giampaolo}, \citenamefont {Morsch}, \citenamefont
  {Giovannetti},\ and\ \citenamefont {Franchini}}]{Catalano2024Frustrating}%
  \BibitemOpen
  \bibfield  {author} {\bibinfo {author} {\bibfnamefont {A.~G.}\ \bibnamefont
  {Catalano}}, \bibinfo {author} {\bibfnamefont {S.~M.}\ \bibnamefont
  {Giampaolo}}, \bibinfo {author} {\bibfnamefont {O.}~\bibnamefont {Morsch}},
  \bibinfo {author} {\bibfnamefont {V.}~\bibnamefont {Giovannetti}},\ and\
  \bibinfo {author} {\bibfnamefont {F.}~\bibnamefont {Franchini}},\ }\bibfield
  {title} {\bibinfo {title} {Frustrating quantum batteries},\ }\href
  {https://doi.org/10.1103/PRXQuantum.5.030319} {\bibfield  {journal} {\bibinfo
   {journal} {PRX Quantum}\ }\textbf {\bibinfo {volume} {5}},\ \bibinfo {pages}
  {030319} (\bibinfo {year} {2024})}\BibitemShut {NoStop}%
\bibitem [{\citenamefont {Razzoli}\ \emph {et~al.}(2025)\citenamefont
  {Razzoli}, \citenamefont {Gemme}, \citenamefont {Khomchenko}, \citenamefont
  {Sassetti}, \citenamefont {Ouerdane}, \citenamefont {Ferraro},\ and\
  \citenamefont {Benenti}}]{Razzoli2025}%
  \BibitemOpen
  \bibfield  {author} {\bibinfo {author} {\bibfnamefont {L.}~\bibnamefont
  {Razzoli}}, \bibinfo {author} {\bibfnamefont {G.}~\bibnamefont {Gemme}},
  \bibinfo {author} {\bibfnamefont {I.}~\bibnamefont {Khomchenko}}, \bibinfo
  {author} {\bibfnamefont {M.}~\bibnamefont {Sassetti}}, \bibinfo {author}
  {\bibfnamefont {H.}~\bibnamefont {Ouerdane}}, \bibinfo {author}
  {\bibfnamefont {D.}~\bibnamefont {Ferraro}},\ and\ \bibinfo {author}
  {\bibfnamefont {G.}~\bibnamefont {Benenti}},\ }\bibfield  {title} {\bibinfo
  {title} {Cyclic solid-state quantum battery: Thermodynamic characterization
  and quantum hardware simulation},\ }\href
  {https://doi.org/10.1088/2058-9565/ad9ed4} {\bibfield  {journal} {\bibinfo
  {journal} {Quantum Sci. Technol.}\ }\textbf {\bibinfo {volume} {10}},\
  \bibinfo {pages} {015064} (\bibinfo {year} {2025})}\BibitemShut {NoStop}%
\bibitem [{\citenamefont {Francica}\ \emph {et~al.}(2017)\citenamefont
  {Francica}, \citenamefont {Goold}, \citenamefont {Plastina},\ and\
  \citenamefont {Paternostro}}]{Francica2017}%
  \BibitemOpen
  \bibfield  {author} {\bibinfo {author} {\bibfnamefont {G.}~\bibnamefont
  {Francica}}, \bibinfo {author} {\bibfnamefont {J.}~\bibnamefont {Goold}},
  \bibinfo {author} {\bibfnamefont {F.}~\bibnamefont {Plastina}},\ and\
  \bibinfo {author} {\bibfnamefont {M.}~\bibnamefont {Paternostro}},\
  }\bibfield  {title} {\bibinfo {title} {Daemonic ergotropy: Enhanced work
  extraction from quantum correlations},\ }\href
  {https://doi.org/10.1038/s41534-017-0012-8} {\bibfield  {journal} {\bibinfo
  {journal} {npj Quantum Inf.}\ }\textbf {\bibinfo {volume} {3}},\ \bibinfo
  {pages} {12} (\bibinfo {year} {2017})}\BibitemShut {NoStop}%
\bibitem [{\citenamefont {Elyasi}\ \emph {et~al.}(2025)\citenamefont {Elyasi},
  \citenamefont {Rossi},\ and\ \citenamefont {Genoni}}]{Elyasi2025}%
  \BibitemOpen
  \bibfield  {author} {\bibinfo {author} {\bibfnamefont {S.~N.}\ \bibnamefont
  {Elyasi}}, \bibinfo {author} {\bibfnamefont {M.~A.~C.}\ \bibnamefont
  {Rossi}},\ and\ \bibinfo {author} {\bibfnamefont {M.~G.}\ \bibnamefont
  {Genoni}},\ }\bibfield  {title} {\bibinfo {title} {Experimental simulation of
  daemonic work extraction in open quantum batteries on a digital quantum
  computer},\ }\href {https://doi.org/10.1088/2058-9565/adae2d} {\bibfield
  {journal} {\bibinfo  {journal} {Quantum Sci. Technol.}\ }\textbf {\bibinfo
  {volume} {10}},\ \bibinfo {pages} {025017} (\bibinfo {year}
  {2025})}\BibitemShut {NoStop}%
\bibitem [{\citenamefont {Liu}\ \emph {et~al.}(2025)\citenamefont {Liu},
  \citenamefont {Zhang}, \citenamefont {Shi}, \citenamefont {Liu},
  \citenamefont {Lu}, \citenamefont {Wang}, \citenamefont {Li}, \citenamefont
  {Li}, \citenamefont {Deng}, \citenamefont {Zhou}, \citenamefont {Liu},
  \citenamefont {Zhang}, \citenamefont {Liang}, \citenamefont {Mei},
  \citenamefont {Ma}, \citenamefont {Liu}, \citenamefont {Liu}, \citenamefont
  {Chen}, \citenamefont {Huang}, \citenamefont {Song}, \citenamefont {Zhao},
  \citenamefont {Tian}, \citenamefont {Xiang}, \citenamefont {Zheng},
  \citenamefont {Nori}, \citenamefont {Xu},\ and\ \citenamefont
  {Fan}}]{Liu2025Processor}%
  \BibitemOpen
  \bibfield  {author} {\bibinfo {author} {\bibfnamefont {Y.}~\bibnamefont
  {Liu}}, \bibinfo {author} {\bibfnamefont {Y.-R.}\ \bibnamefont {Zhang}},
  \bibinfo {author} {\bibfnamefont {Y.-H.}\ \bibnamefont {Shi}}, \bibinfo
  {author} {\bibfnamefont {T.}~\bibnamefont {Liu}}, \bibinfo {author}
  {\bibfnamefont {C.}~\bibnamefont {Lu}}, \bibinfo {author} {\bibfnamefont
  {Y.-Y.}\ \bibnamefont {Wang}}, \bibinfo {author} {\bibfnamefont
  {H.}~\bibnamefont {Li}}, \bibinfo {author} {\bibfnamefont {T.-M.}\
  \bibnamefont {Li}}, \bibinfo {author} {\bibfnamefont {C.-L.}\ \bibnamefont
  {Deng}}, \bibinfo {author} {\bibfnamefont {S.-Y.}\ \bibnamefont {Zhou}},
  \bibinfo {author} {\bibfnamefont {T.}~\bibnamefont {Liu}}, \bibinfo {author}
  {\bibfnamefont {J.-C.}\ \bibnamefont {Zhang}}, \bibinfo {author}
  {\bibfnamefont {G.-H.}\ \bibnamefont {Liang}}, \bibinfo {author}
  {\bibfnamefont {Z.-Y.}\ \bibnamefont {Mei}}, \bibinfo {author} {\bibfnamefont
  {W.-G.}\ \bibnamefont {Ma}}, \bibinfo {author} {\bibfnamefont {H.-T.}\
  \bibnamefont {Liu}}, \bibinfo {author} {\bibfnamefont {Z.-H.}\ \bibnamefont
  {Liu}}, \bibinfo {author} {\bibfnamefont {C.-T.}\ \bibnamefont {Chen}},
  \bibinfo {author} {\bibfnamefont {K.}~\bibnamefont {Huang}}, \bibinfo
  {author} {\bibfnamefont {X.}~\bibnamefont {Song}}, \bibinfo {author}
  {\bibfnamefont {S.~P.}\ \bibnamefont {Zhao}}, \bibinfo {author}
  {\bibfnamefont {Y.}~\bibnamefont {Tian}}, \bibinfo {author} {\bibfnamefont
  {Z.}~\bibnamefont {Xiang}}, \bibinfo {author} {\bibfnamefont
  {D.}~\bibnamefont {Zheng}}, \bibinfo {author} {\bibfnamefont
  {F.}~\bibnamefont {Nori}}, \bibinfo {author} {\bibfnamefont {K.}~\bibnamefont
  {Xu}},\ and\ \bibinfo {author} {\bibfnamefont {H.}~\bibnamefont {Fan}},\
  }\bibfield  {title} {\bibinfo {title} {Interplay between disorder and
  topology in {Thouless} pumping on a superconducting quantum processor},\
  }\href {https://doi.org/10.1038/s41467-024-55343-2} {\bibfield  {journal}
  {\bibinfo  {journal} {Nat. Commun.}\ }\textbf {\bibinfo {volume} {16}},\
  \bibinfo {pages} {108} (\bibinfo {year} {2025})}\BibitemShut {NoStop}%
\bibitem [{Sup()}]{Supplemental}%
  \BibitemOpen
  \href@noop {} {}\bibinfo {note} {See the Supplemental Material appended to this
  manuscript for model definitions, derivations, the full
  transition sweep, robustness data, and additional controls}\BibitemShut
  {NoStop}%
\bibitem [{\citenamefont {Ciccarello}\ \emph {et~al.}(2022)\citenamefont
  {Ciccarello}, \citenamefont {Lorenzo}, \citenamefont {Giovannetti},\ and\
  \citenamefont {Palma}}]{Ciccarello2022}%
  \BibitemOpen
  \bibfield  {author} {\bibinfo {author} {\bibfnamefont {F.}~\bibnamefont
  {Ciccarello}}, \bibinfo {author} {\bibfnamefont {S.}~\bibnamefont {Lorenzo}},
  \bibinfo {author} {\bibfnamefont {V.}~\bibnamefont {Giovannetti}},\ and\
  \bibinfo {author} {\bibfnamefont {G.~M.}\ \bibnamefont {Palma}},\ }\bibfield
  {title} {\bibinfo {title} {Quantum collision models: Open system dynamics
  from repeated interactions},\ }\href
  {https://doi.org/10.1016/j.physrep.2022.01.001} {\bibfield  {journal}
  {\bibinfo  {journal} {Phys. Rep.}\ }\textbf {\bibinfo {volume} {954}},\
  \bibinfo {pages} {1} (\bibinfo {year} {2022})}\BibitemShut {NoStop}%
\bibitem [{\citenamefont {Brown}\ \emph {et~al.}(2016)\citenamefont {Brown},
  \citenamefont {Friis},\ and\ \citenamefont {Huber}}]{Brown2016}%
  \BibitemOpen
  \bibfield  {author} {\bibinfo {author} {\bibfnamefont {E.~G.}\ \bibnamefont
  {Brown}}, \bibinfo {author} {\bibfnamefont {N.}~\bibnamefont {Friis}},\ and\
  \bibinfo {author} {\bibfnamefont {M.}~\bibnamefont {Huber}},\ }\bibfield
  {title} {\bibinfo {title} {Passivity and practical work extraction using
  {Gaussian} operations},\ }\href
  {https://doi.org/10.1088/1367-2630/18/11/113028} {\bibfield  {journal}
  {\bibinfo  {journal} {New J. Phys.}\ }\textbf {\bibinfo {volume} {18}},\
  \bibinfo {pages} {113028} (\bibinfo {year} {2016})}\BibitemShut {NoStop}%
\bibitem [{\citenamefont {Strasberg}\ \emph {et~al.}(2017)\citenamefont
  {Strasberg}, \citenamefont {Schaller}, \citenamefont {Brandes},\ and\
  \citenamefont {Esposito}}]{Strasberg2017}%
  \BibitemOpen
  \bibfield  {author} {\bibinfo {author} {\bibfnamefont {P.}~\bibnamefont
  {Strasberg}}, \bibinfo {author} {\bibfnamefont {G.}~\bibnamefont {Schaller}},
  \bibinfo {author} {\bibfnamefont {T.}~\bibnamefont {Brandes}},\ and\ \bibinfo
  {author} {\bibfnamefont {M.}~\bibnamefont {Esposito}},\ }\bibfield  {title}
  {\bibinfo {title} {Quantum and information thermodynamics: A unifying
  framework based on repeated interactions},\ }\href
  {https://doi.org/10.1103/PhysRevX.7.021003} {\bibfield  {journal} {\bibinfo
  {journal} {Phys. Rev. X}\ }\textbf {\bibinfo {volume} {7}},\ \bibinfo {pages}
  {021003} (\bibinfo {year} {2017})}\BibitemShut {NoStop}%
\bibitem [{\citenamefont {Peschel}(2003)}]{Peschel2003}%
  \BibitemOpen
  \bibfield  {author} {\bibinfo {author} {\bibfnamefont {I.}~\bibnamefont
  {Peschel}},\ }\bibfield  {title} {\bibinfo {title} {Calculation of reduced
  density matrices from correlation functions},\ }\href
  {https://doi.org/10.1088/0305-4470/36/14/101} {\bibfield  {journal} {\bibinfo
   {journal} {J. Phys. A}\ }\textbf {\bibinfo {volume} {36}},\ \bibinfo {pages}
  {L205} (\bibinfo {year} {2003})}\BibitemShut {NoStop}%
\bibitem [{\citenamefont {Wilson}(1927)}]{Wilson1927}%
  \BibitemOpen
  \bibfield  {author} {\bibinfo {author} {\bibfnamefont {E.~B.}\ \bibnamefont
  {Wilson}},\ }\bibfield  {title} {\bibinfo {title} {Probable inference, the
  law of succession, and statistical inference},\ }\href
  {https://doi.org/10.1080/01621459.1927.10502953} {\bibfield  {journal}
  {\bibinfo  {journal} {J. Am. Stat. Assoc.}\ }\textbf {\bibinfo {volume}
  {22}},\ \bibinfo {pages} {209} (\bibinfo {year} {1927})}\BibitemShut
  {NoStop}%
\bibitem [{\citenamefont {Andolina}\ \emph {et~al.}(2019)\citenamefont
  {Andolina}, \citenamefont {Keck}, \citenamefont {Mari}, \citenamefont
  {Campisi}, \citenamefont {Giovannetti},\ and\ \citenamefont
  {Polini}}]{Andolina2019}%
  \BibitemOpen
  \bibfield  {author} {\bibinfo {author} {\bibfnamefont {G.~M.}\ \bibnamefont
  {Andolina}}, \bibinfo {author} {\bibfnamefont {M.}~\bibnamefont {Keck}},
  \bibinfo {author} {\bibfnamefont {A.}~\bibnamefont {Mari}}, \bibinfo {author}
  {\bibfnamefont {M.}~\bibnamefont {Campisi}}, \bibinfo {author} {\bibfnamefont
  {V.}~\bibnamefont {Giovannetti}},\ and\ \bibinfo {author} {\bibfnamefont
  {M.}~\bibnamefont {Polini}},\ }\bibfield  {title} {\bibinfo {title}
  {Extractable work, the role of correlations, and asymptotic freedom in
  quantum batteries},\ }\href {https://doi.org/10.1103/PhysRevLett.122.047702}
  {\bibfield  {journal} {\bibinfo  {journal} {Phys. Rev. Lett.}\ }\textbf
  {\bibinfo {volume} {122}},\ \bibinfo {pages} {047702} (\bibinfo {year}
  {2019})}\BibitemShut {NoStop}%
\end{thebibliography}%


\begin{thebibliography}{6}%
\makeatletter
\providecommand \@ifxundefined [1]{%
 \@ifx{#1\undefined}
}%
\providecommand \@ifnum [1]{%
 \ifnum #1\expandafter \@firstoftwo
 \else \expandafter \@secondoftwo
 \fi
}%
\providecommand \@ifx [1]{%
 \ifx #1\expandafter \@firstoftwo
 \else \expandafter \@secondoftwo
 \fi
}%
\providecommand \natexlab [1]{#1}%
\providecommand \enquote  [1]{``#1''}%
\providecommand \bibnamefont  [1]{#1}%
\providecommand \bibfnamefont [1]{#1}%
\providecommand \citenamefont [1]{#1}%
\providecommand \href@noop [0]{\@secondoftwo}%
\providecommand \href [0]{\begingroup \@sanitize@url \@href}%
\providecommand \@href[1]{\@@startlink{#1}\@@href}%
\providecommand \@@href[1]{\endgroup#1\@@endlink}%
\providecommand \@sanitize@url [0]{\catcode `\\12\catcode `\$12\catcode
  `\&12\catcode `\#12\catcode `\^12\catcode `\_12\catcode `\%12\relax}%
\providecommand \@@startlink[1]{}%
\providecommand \@@endlink[0]{}%
\providecommand \url  [0]{\begingroup\@sanitize@url \@url }%
\providecommand \@url [1]{\endgroup\@href {#1}{\urlprefix }}%
\providecommand \urlprefix  [0]{URL }%
\providecommand \Eprint [0]{\href }%
\providecommand \doibase [0]{https://doi.org/}%
\providecommand \selectlanguage [0]{\@gobble}%
\providecommand \bibinfo  [0]{\@secondoftwo}%
\providecommand \bibfield  [0]{\@secondoftwo}%
\providecommand \translation [1]{[#1]}%
\providecommand \BibitemOpen [0]{}%
\providecommand \bibitemStop [0]{}%
\providecommand \bibitemNoStop [0]{.\EOS\space}%
\providecommand \EOS [0]{\spacefactor3000\relax}%
\providecommand \BibitemShut  [1]{\csname bibitem#1\endcsname}%
\let\auto@bib@innerbib\@empty
\bibitem [{\citenamefont {Liu}\ \emph {et~al.}(2025)\citenamefont {Liu},
  \citenamefont {Zhang}, \citenamefont {Shi}, \citenamefont {Liu},
  \citenamefont {Lu}, \citenamefont {Wang}, \citenamefont {Li}, \citenamefont
  {Li}, \citenamefont {Deng}, \citenamefont {Zhou}, \citenamefont {Liu},
  \citenamefont {Zhang}, \citenamefont {Liang}, \citenamefont {Mei},
  \citenamefont {Ma}, \citenamefont {Liu}, \citenamefont {Liu}, \citenamefont
  {Chen}, \citenamefont {Huang}, \citenamefont {Song}, \citenamefont {Zhao},
  \citenamefont {Tian}, \citenamefont {Xiang}, \citenamefont {Zheng},
  \citenamefont {Nori}, \citenamefont {Xu},\ and\ \citenamefont
  {Fan}}]{Liu2025Processor}%
  \BibitemOpen
  \bibfield  {author} {\bibinfo {author} {\bibfnamefont {Y.}~\bibnamefont
  {Liu}}, \bibinfo {author} {\bibfnamefont {Y.-R.}\ \bibnamefont {Zhang}},
  \bibinfo {author} {\bibfnamefont {Y.-H.}\ \bibnamefont {Shi}}, \bibinfo
  {author} {\bibfnamefont {T.}~\bibnamefont {Liu}}, \bibinfo {author}
  {\bibfnamefont {C.}~\bibnamefont {Lu}}, \bibinfo {author} {\bibfnamefont
  {Y.-Y.}\ \bibnamefont {Wang}}, \bibinfo {author} {\bibfnamefont
  {H.}~\bibnamefont {Li}}, \bibinfo {author} {\bibfnamefont {T.-M.}\
  \bibnamefont {Li}}, \bibinfo {author} {\bibfnamefont {C.-L.}\ \bibnamefont
  {Deng}}, \bibinfo {author} {\bibfnamefont {S.-Y.}\ \bibnamefont {Zhou}},
  \bibinfo {author} {\bibfnamefont {T.}~\bibnamefont {Liu}}, \bibinfo {author}
  {\bibfnamefont {J.-C.}\ \bibnamefont {Zhang}}, \bibinfo {author}
  {\bibfnamefont {G.-H.}\ \bibnamefont {Liang}}, \bibinfo {author}
  {\bibfnamefont {Z.-Y.}\ \bibnamefont {Mei}}, \bibinfo {author} {\bibfnamefont
  {W.-G.}\ \bibnamefont {Ma}}, \bibinfo {author} {\bibfnamefont {H.-T.}\
  \bibnamefont {Liu}}, \bibinfo {author} {\bibfnamefont {Z.-H.}\ \bibnamefont
  {Liu}}, \bibinfo {author} {\bibfnamefont {C.-T.}\ \bibnamefont {Chen}},
  \bibinfo {author} {\bibfnamefont {K.}~\bibnamefont {Huang}}, \bibinfo
  {author} {\bibfnamefont {X.}~\bibnamefont {Song}}, \bibinfo {author}
  {\bibfnamefont {S.~P.}\ \bibnamefont {Zhao}}, \bibinfo {author}
  {\bibfnamefont {Y.}~\bibnamefont {Tian}}, \bibinfo {author} {\bibfnamefont
  {Z.}~\bibnamefont {Xiang}}, \bibinfo {author} {\bibfnamefont
  {D.}~\bibnamefont {Zheng}}, \bibinfo {author} {\bibfnamefont
  {F.}~\bibnamefont {Nori}}, \bibinfo {author} {\bibfnamefont {K.}~\bibnamefont
  {Xu}},\ and\ \bibinfo {author} {\bibfnamefont {H.}~\bibnamefont {Fan}},\
  }\bibfield  {title} {\bibinfo {title} {Interplay between disorder and
  topology in {Thouless} pumping on a superconducting quantum processor},\
  }\href {https://doi.org/10.1038/s41467-024-55343-2} {\bibfield  {journal}
  {\bibinfo  {journal} {Nat. Commun.}\ }\textbf {\bibinfo {volume} {16}},\
  \bibinfo {pages} {108} (\bibinfo {year} {2025})}\BibitemShut {NoStop}%
\bibitem [{\citenamefont {Allahverdyan}\ \emph {et~al.}(2004)\citenamefont
  {Allahverdyan}, \citenamefont {Balian},\ and\ \citenamefont {{Th. M.
  Nieuwenhuizen}}}]{Allahverdyan2004}%
  \BibitemOpen
  \bibfield  {author} {\bibinfo {author} {\bibfnamefont {A.~E.}\ \bibnamefont
  {Allahverdyan}}, \bibinfo {author} {\bibfnamefont {R.}~\bibnamefont
  {Balian}},\ and\ \bibinfo {author} {\bibnamefont {{Th. M. Nieuwenhuizen}}},\
  }\bibfield  {title} {\bibinfo {title} {Maximal work extraction from finite
  quantum systems},\ }\href {https://doi.org/10.1209/epl/i2004-10101-2}
  {\bibfield  {journal} {\bibinfo  {journal} {Europhys. Lett.}\ }\textbf
  {\bibinfo {volume} {67}},\ \bibinfo {pages} {565} (\bibinfo {year}
  {2004})}\BibitemShut {NoStop}%
\bibitem [{\citenamefont {Brown}\ \emph {et~al.}(2016)\citenamefont {Brown},
  \citenamefont {Friis},\ and\ \citenamefont {Huber}}]{Brown2016}%
  \BibitemOpen
  \bibfield  {author} {\bibinfo {author} {\bibfnamefont {E.~G.}\ \bibnamefont
  {Brown}}, \bibinfo {author} {\bibfnamefont {N.}~\bibnamefont {Friis}},\ and\
  \bibinfo {author} {\bibfnamefont {M.}~\bibnamefont {Huber}},\ }\bibfield
  {title} {\bibinfo {title} {Passivity and practical work extraction using
  {Gaussian} operations},\ }\href
  {https://doi.org/10.1088/1367-2630/18/11/113028} {\bibfield  {journal}
  {\bibinfo  {journal} {New J. Phys.}\ }\textbf {\bibinfo {volume} {18}},\
  \bibinfo {pages} {113028} (\bibinfo {year} {2016})}\BibitemShut {NoStop}%
\bibitem [{\citenamefont {Peschel}(2003)}]{Peschel2003}%
  \BibitemOpen
  \bibfield  {author} {\bibinfo {author} {\bibfnamefont {I.}~\bibnamefont
  {Peschel}},\ }\bibfield  {title} {\bibinfo {title} {Calculation of reduced
  density matrices from correlation functions},\ }\href
  {https://doi.org/10.1088/0305-4470/36/14/101} {\bibfield  {journal} {\bibinfo
   {journal} {J. Phys. A}\ }\textbf {\bibinfo {volume} {36}},\ \bibinfo {pages}
  {L205} (\bibinfo {year} {2003})}\BibitemShut {NoStop}%
\bibitem [{\citenamefont {Strasberg}\ \emph {et~al.}(2017)\citenamefont
  {Strasberg}, \citenamefont {Schaller}, \citenamefont {Brandes},\ and\
  \citenamefont {Esposito}}]{Strasberg2017}%
  \BibitemOpen
  \bibfield  {author} {\bibinfo {author} {\bibfnamefont {P.}~\bibnamefont
  {Strasberg}}, \bibinfo {author} {\bibfnamefont {G.}~\bibnamefont {Schaller}},
  \bibinfo {author} {\bibfnamefont {T.}~\bibnamefont {Brandes}},\ and\ \bibinfo
  {author} {\bibfnamefont {M.}~\bibnamefont {Esposito}},\ }\bibfield  {title}
  {\bibinfo {title} {Quantum and information thermodynamics: A unifying
  framework based on repeated interactions},\ }\href
  {https://doi.org/10.1103/PhysRevX.7.021003} {\bibfield  {journal} {\bibinfo
  {journal} {Phys. Rev. X}\ }\textbf {\bibinfo {volume} {7}},\ \bibinfo {pages}
  {021003} (\bibinfo {year} {2017})}\BibitemShut {NoStop}%
\bibitem [{\citenamefont {Wilson}(1927)}]{Wilson1927}%
  \BibitemOpen
  \bibfield  {author} {\bibinfo {author} {\bibfnamefont {E.~B.}\ \bibnamefont
  {Wilson}},\ }\bibfield  {title} {\bibinfo {title} {Probable inference, the
  law of succession, and statistical inference},\ }\href
  {https://doi.org/10.1080/01621459.1927.10502953} {\bibfield  {journal}
  {\bibinfo  {journal} {J. Am. Stat. Assoc.}\ }\textbf {\bibinfo {volume}
  {22}},\ \bibinfo {pages} {209} (\bibinfo {year} {1927})}\BibitemShut
  {NoStop}%
\end{thebibliography}%

\end{document}


\raggedbottom
\renewcommand{\theequation}{S\arabic{equation}}
\renewcommand{\thefigure}{S\arabic{figure}}
\renewcommand{\thetable}{S\Roman{table}}

\title{Supplemental Material for ``Boundary-accessible work in a collision-charged topological quantum battery''}
\author{Osman Or\c{c}un Keskin}
\affiliation{Department of Physics, Ko\c{c} University, 34450 Sar\i yer, Istanbul, T\"urkiye}
\author{\"Ozg\"ur E. M\"ustecapl{\i}o\u{g}lu}
\affiliation{Department of Physics, Ko\c{c} University, 34450 Sar\i yer, Istanbul, T\"urkiye}
\affiliation{T\"UB\.ITAK Research Institute for Fundamental Sciences (TBAE), 41470 Gebze, Kocaeli, T\"urkiye}
\date{\today}
\maketitle

This document specifies the numerical protocols and derives the natural-occupation and passive-state constructions used in the main manuscript. It also presents the full fixed-control sweep across the SSH transition, baseline and finite-size controls, and selected noise and disorder checks.

\section{Complete model specification}

The battery is a hard-core SSH chain of $N$ unit cells and $2N$ sites,
\begin{align}
H_B={}&\epsilon_0\sum_{j=1}^{N}(n_{A_j}+n_{B_j})
+v\sum_{j=1}^{N}(\sigma_{A_j}^{+}\sigma_{B_j}^{-}+\mathrm{H.c.})\nonumber\\
&+w\sum_{j=1}^{N-1}(\sigma_{B_j}^{+}\sigma_{A_{j+1}}^{-}+\mathrm{H.c.}).
\label{eqS:HB}
\end{align}
Energies are expressed in units of the hopping scale $J$, defined by $v+w=2J$, and times in units of $J^{-1}$; figure axes, legends, and Table~\ref{tabS:parameters} quote numerical values in these units. The empty state is the unique ground state because $\epsilon_0=2.5J>|v|+|w|=2J$. The one-particle eigenvectors and winding classification are unchanged by this uniform shift. The baseline phases are $(v,w)=(1.5,0.5)J$ and $(0.5,1.5)J$ for the trivial and topological chains, respectively. The baseline detuning $\Delta_A=\omega_A-\epsilon_0=J$ coincides with the lower edge of the upper bulk band in both phases. A fresh excited ancilla with $H_A=\omega_An_A$ collides with site $\ell$ through
\begin{equation}
U_\ell=\exp[-i\tau(H_A+H_B+g\sigma_A^+\sigma_\ell^-+g\sigma_A^-\sigma_\ell^+)].
\label{eqS:collision}
\end{equation}
The battery begins in vacuum. Edge injection uses $\ell=A_1$ and the bulk control uses $\ell=A_2$. For the Jordan--Wigner modes $c_i$, we write $(C_B)_{ij}=\langle c_j^\dagger c_i\rangle$, $N_B=\Tr C_B$, and $C_A=[p_e]$ for an incoming ancilla with excited-state population $p_e$. Table~\ref{tabS:parameters} summarizes the parameter sets.

All static-disorder and collision-control ensembles use independent variables uniformly distributed on $[-1,1]$. Hopping disorder acts as
\begin{equation}
v_j=v(1+s_hx_j),\qquad w_j=w(1+s_hy_j),
\end{equation}
and on-site disorder as $\epsilon_j=\epsilon_0+s_\epsilon z_j$, with independent $x_j,y_j,z_j\sim\mathcal U[-1,1]$. Charger-frequency and collision-duration noise are $\omega_{A,m}=\omega_A+s_\omega x_m$ and $\tau_m=\tau(1+s_\tau y_m)$, respectively. Thus $s_h$ and $s_\tau$ are relative amplitudes, while $s_\epsilon$ and $s_\omega$ are energy half-widths reported in units of $J$. The scans use
\begin{align}
s_h&\in\{0,0.025,0.05,0.075,0.10,0.15,0.20\},\nonumber\\
s_\epsilon/J,s_\omega/J&\in\{0,0.05,0.10,0.15,0.20,0.25,0.30\},\nonumber\\
s_\tau&\in\{0,0.025,0.05,0.075,0.10,0.15,0.20\},
\end{align}
with 50 realizations per point. The paired load-disorder calculation uses $s_h\in\{0,0.05,0.10,0.15,0.20\}$ and $s_\epsilon/J\in\{0,0.15,0.30,0.45,0.60\}$, with the same normalized draws applied to the two SSH phases in each of 200 pairs. The base pseudorandom seed is 3902026; independent ensemble families use fixed integer offsets from this seed. Curves show ensemble means, and bands show one sample standard deviation.

For a channel interval $\Delta t$, let $q=e^{-\gamma\Delta t}$. Local qubit dephasing during the small-$N$ charging calculations uses $K_0=\sqrt{(1+q)/2}\,I$ and $K_1=\sqrt{(1-q)/2}\,\sigma_z$, while local qubit amplitude loss uses $K_0=|0\rangle\langle0|+\sqrt q\,|1\rangle\langle1|$ and $K_1=\sqrt{1-q}\,\sigma^-$. During large-$N$ extraction, the same rate convention gives $C\mapsto qC$ for fermionic-mode loss and $(C)_{ij}\mapsto q^2(C)_{ij}$ for $i\ne j$ under independent dephasing. Both charging- and extraction-noise scans use $\gamma/J\in\{0,0.005,0.010,0.015,0.025,0.050,0.075,0.100,0.125,0.150\}$.

\begin{table}[!htbp]
\caption{Numerical parameter sets. Energies are given in units of $J$ and times in units of $J^{-1}$; parameters absent from a specialized row retain their baseline values.}
\label{tabS:parameters}
\begin{ruledtabular}
\begin{tabular}{p{0.24\textwidth}p{0.70\textwidth}}
Calculation & Parameters \\
\hline
Baseline exact & $N=3$, $\epsilon_0=2.5$, $\omega_A=3.5$, $g=0.3$, $\tau=2$, six collisions; vacuum battery and excited fresh ancillas. \\
SSH phases & Trivial $(v,w)=(1.5,0.5)$; topological $(v,w)=(0.5,1.5)$. \\
Finite size & Edge injection; fixed control for $N=4,\ldots,30$; detuning/coupling maps through $N=100$; collision-horizon map through $N=30$ and $n=10^6$. \\
Steady-state horizon & $N=6$; collision counts through $10^7$; $99\%$ filling at $429{,}774$ collisions (trivial) and $1{,}341{,}872$ collisions (topological). \\
Long chain & Exact covariance method, $N=100$, 31 detunings $\Delta_A=\omega_A-\epsilon_0\in[-1.5,1.5]$. \\
Dimerization sweep & $N=100$, $\delta\in[-0.75,0.75]$ in 151 steps, $v=J(1-\delta)$, $w=J(1+\delta)$; six resonant collisions and one common load pulse with $g_L=0.3$, $\omega_L=\epsilon_0$, and $t_L=5.6$. \\
Passivity & Natural occupations at $\Delta_A=0,1$; $\epsilon_0\in[2.05,10]$ at fixed $\Delta_A=1$; crossover controls at $N=3,4,6,20,100$. \\
Local discharge & Empty load at $A_1$; $N=3,4,6,20,100$; $\Delta_A=0,1$; $g_L=0.3$, $\omega_L=\epsilon_0$; 101 load times on $[0,10]$ and the same pulse in both phases. \\
Cycle & $N=20$; $n=1,\ldots,20$; $\Delta_A=0,1$; $t_{\mathrm{oh}}=0,2,5,10$. \\
Extraction noise & $N=20$; ten dephasing or fermionic-mode loss rates through $0.15$; channel active on all battery fermionic modes and the load during an 80-step symmetric propagation. \\
Control noise & $N=20$; 50 seeded samples per charger-frequency and collision-duration noise point. \\
Long-contact disorder spectroscopy & $N=20$; six collisions; $(g,\tau)=(0.05J,20J^{-1})$; 25 detunings on $[-0.3J,0.3J]$; 200 independent realizations per phase and disorder class. \\
Paired load disorder & $N=20$; 200 matched realizations at each of five hopping or on-site strengths; fixed charge and load controls. \\
Noisy charging & Full-density $N=3$; ten local-dephasing or amplitude-loss rates through $0.15$, followed by an ideal load pulse. \\
\end{tabular}
\end{ruledtabular}
\end{table}

\subsection{Illustrative hardware scale}

For scale, choosing $J/(2\pi)=2$ MHz, compatible with a published frequency-tunable transmon processor, maps $g=0.3J$ to $g/(2\pi)=0.6$ MHz, $\tau=2J^{-1}$ to $159$ ns, and $t_L=5.6J^{-1}$ to $446$ ns \cite{Liu2025Processor}. Six collisions followed by the load contact then take approximately $1.40~\mu$s, compared with the processor's reported mean $T_1=21.0~\mu$s (Supplementary Table~1 of Ref.~\cite{Liu2025Processor}). A uniform carrier can be removed in a rotating-frame implementation, but the absolute energies and global ergotropies quoted in this work refer to $H_B$ as written with $\epsilon_0=2.5J$. The activation criterion and the ratio $p_L/\nu_{\max}$ do not depend on this common energy choice.

This conversion provides an illustrative physical timescale for the dimensionless parameters. Device-level modeling would additionally incorporate pulse shapes, crosstalk, leakage, readout, reset, and calibrated noise.

\section{Work measures and exact constructions}

\subsection{Global and quadratic ergotropy}

Let $r_k$ be the eigenvalues of $\rho_B$ in nonincreasing order and $\varepsilon_k$ the many-body energies in nondecreasing order. The globally passive energy and ergotropy are \cite{Allahverdyan2004}
\begin{equation}
E_{\mathrm{pas}}=\sum_k r_k\varepsilon_k,\qquad
\W=\Tr(\rho_BH_B)-E_{\mathrm{pas}}.
\label{eqS:global}
\end{equation}
For any battery state, the eigenvectors of $C_B$ are its natural orbitals and the eigenvalues $\nu_j$ are its natural occupations. Here ``quadratic'' means number-conserving quadratic evolution in the Jordan--Wigner fermionic modes. Such unitaries act as $C_B\mapsto uC_Bu^\dagger$, while a quadratic battery Hamiltonian has energy $E_B=\Tr(hC_B)$. The trace inequality
\begin{equation}
\min_u\Tr(h\,uC_Bu^\dagger)=\sum_j\nu_j^\downarrow e_j^\uparrow
\end{equation}
therefore gives, for any battery state,
\begin{equation}
E_{\mathrm{pas}}^{(2)}=\sum_j\nu_j^\downarrow e_j^\uparrow,\qquad
\W^{(2)}=E_B-E_{\mathrm{pas}}^{(2)}.
\label{eqS:quadratic}
\end{equation}
Reference~\cite{Brown2016} gives the analogous construction for bosonic Gaussian operations.
Writing $e_j=\epsilon_0+\widetilde e_j$ yields
\begin{equation}
E_{\mathrm{pas}}^{(2)}=\epsilon_0N_B+\sum_j\nu_j^\downarrow\widetilde e_j^\uparrow,
\end{equation}
so $\W^{(2)}$ is exactly invariant under a common shift of $\epsilon_0$, $\omega_A$, and $\omega_L$ at fixed detunings. The shift multiplies each number-conserving propagator by a phase generated by the conserved total number and leaves $C_B$ unchanged. Hence $\nu_{\max}$, $p_L$, the load-activation criterion, and $\W^{(2)}$ are independent of this common shift. Global ergotropy need not be invariant because unrestricted work-extraction unitaries can change excitation number, and the many-body passive ordering then depends on the absolute energy offset.

\subsection{Natural-occupation load bound}

Before extraction, the load is empty and uncorrelated, so the joint one-body correlation matrix is
\begin{equation}
C_{BL}^{(0)}=0_L\oplus C_B.
\end{equation}
A number-conserving quadratic unitary is represented by a one-body unitary $u$. If $|L\rangle$ denotes the output load mode, its final population is
\begin{equation}
p_L=\langle L|uC_{BL}^{(0)}u^\dagger|L\rangle
=\bm u_B^\dagger C_B\bm u_B,
\end{equation}
where $\bm u_B=P_Bu^\dagger|L\rangle$, $P_B$ projects onto the battery block, and $\|\bm u_B\|\leq1$. The Rayleigh--Ritz theorem gives
\begin{equation}
p_L\leq\nu_{\max}(C_B).
\label{eqS:bound}
\end{equation}
Equation~\eqref{eqS:bound} is a population ceiling that depends only on $C_B$ and applies to Gaussian and non-Gaussian battery states. It is attained by the one-body unitary that maps the dominant natural orbital onto the load mode, so the bound is tight within the quadratic class. The charged battery is number-diagonal, $[\rho_B,\widehat N_B]=0$; the load starts in $|0\rangle_L$, and $[U_{BL},\widehat N_B+n_L]=0$. The final joint state is therefore number-diagonal, so its one-mode load marginal commutes with $n_L$ and equals $\rho_L=(1-p_L)|0\rangle\langle0|+p_L|1\rangle\langle1|$. Hence $E_L=\omega_Lp_L$ and $\W_L=\omega_L\max(0,2p_L-1)$, which gives the work ceiling stated in the main text. We call $\nu_{\max}>1/2$ a natural-orbital inversion: it is the activation condition for an initially empty number-diagonal two-level load under this quadratic extraction class. The implemented pulse is a local boundary coupling, while the general population bound covers arbitrary number-conserving quadratic mode rotations. Here $\W_L$ is the ergotropy available in the decoupled load; harvesting it requires a subsequent cyclic local drive that interchanges its populations when $p_L>1/2$.

\subsection{Dense and covariance solvers}

For $N=3$ and selected $N=4$ controls, all operators are built on the complete hard-core Hilbert space. Each collision is followed by an explicit partial trace over the ancilla, and the many-body battery state is diagonalized directly.

For the memoryless edge geometry, the Jordan--Wigner transformation maps Eq.~\eqref{eqS:HB} to a number-conserving fermion Hamiltonian. With the ancilla placed adjacent to the first battery mode, a one-particle unitary updates
\begin{equation}
C'_B=[U(C_A\oplus C_B)U^\dagger]_B.
\end{equation}
Vacuum input and number-diagonal single-mode ancillas preserve number-diagonal Gaussianity. The many-body eigenvalues of the state are all products $\prod_q\nu_q^{s_q}(1-\nu_q)^{1-s_q}$ \cite{Peschel2003}. Since each fresh ancilla adds at most a rank-one source, $\operatorname{rank}C_n\leq n$. Six collisions therefore leave at most 64 nonzero state eigenvalues at every $N$. The global passive energy is evaluated by pairing those weights with the 64 lowest many-body energy subset sums.

\subsection{Switching identities and statistical intervals}

For $n$ charging collisions, the aggregate charger switching work is
\begin{equation}
W_{\mathrm{sw}}^{(\mathrm{ch})}=\sum_{m=1}^{n}
\left(\Delta E_B^{(m)}+\Delta E_A^{(m)}\right)
=-\sum_{m=1}^{n}\langle H_{\mathrm{int}}\rangle_{m,\tau},
\end{equation}
where the second expression follows from the interaction energy at separation \cite{Strasberg2017}. The load obeys $W_{\mathrm{sw}}^{(L)}=\Delta E_B+E_L=-\langle H_{BL}\rangle_{t_L}$.

At resonance, $\omega_A=\epsilon_0$, the shifted one-body Hamiltonian $H-\epsilon_0\widehat N_{\mathrm{tot}}$ is a real nearest-neighbor hopping matrix on a bipartite chain. The charger is assigned to the sublattice opposite to $A_1$, as is a resonant load coupled at the same boundary. The one-body propagator then has the form $u=DOD^\dagger$, where $O$ is real orthogonal and $D=\Gamma^{1/2}$ for the chiral operator $\Gamma=\operatorname{diag}(I_A,-I_B)$; equivalently, $D=1$ on the $A$ sublattice and $D=i$ on the $B$ sublattice, including the charger and load. Starting from the number-diagonal inputs used here, the reduced battery correlation matrix consequently has the form $C_B=DRD^\dagger$ with $R$ real symmetric after every collision. All nearest-neighbor coherences are purely imaginary, so every bond-energy contribution vanishes and
\begin{equation}
E_B=\epsilon_0N_B,\qquad
\langle H_{\mathrm{int}}\rangle=0,\qquad
W_{\mathrm{sw}}^{(\mathrm{ch})}=0.
\label{eqS:resonance-identities}
\end{equation}
If $\omega_L=\epsilon_0$, the same form is preserved during discharge, giving $\langle H_{BL}\rangle_t=0$ and $W_{\mathrm{sw}}^{(L)}=0$. These identities hold in both SSH phases for any $N$, collision number, coupling, and collision duration. They survive hopping disorder, dephasing, and uniform fermionic-mode loss, whereas on-site disorder and nonzero charger detuning break them. It follows that resonant extraction has $\W_{\mathrm{net}}=\W_L$ and
\begin{equation}
\eta_{\mathrm{cyc}}=\frac{[2p_L-1]_+}{N_B}.
\end{equation}

For the paired fixed-pulse disorder ensemble, phase differences use the same normalized perturbation draw for the trivial and topological Hamiltonians. If $d_i$ is the $i$th work contrast, $\overline d=M^{-1}\sum_i d_i$, and $s_d^2=(M-1)^{-1}\sum_i(d_i-\overline d)^2$, the two-sided mean interval is
\begin{equation}
\overline d\pm t_{0.975,M-1}\,s_d/\sqrt M,
\end{equation}
with $M=200$. Active-load fractions use 95\% Wilson score intervals \cite{Wilson1927}. The ensemble summaries also include the median, fifth and 95th percentiles, and $\Pr(d_i>0)$.

\subsection{Numerical reproducibility}

For $N=2,3$ and both SSH phases, the independently implemented full-density and covariance solvers agree after six baseline collisions in $N_B$, $n_{A_1}$, $E_B$, $E_{\mathrm{pas}}$, and $\W$, with a largest absolute difference of $1.56\times10^{-14}$. The fixed seeds and phase-matched disorder draws specified above define the reported ensembles.

For the extraction-noise scan in main-text Fig.~5(b), we repeated the symmetric split-step propagation with 80, 160, and 320 steps at fixed physical parameters. For local dephasing, the maximum absolute changes over both phases and all sampled noise rates were $8.366\times10^{-7}$ and $2.087\times10^{-7}$ in $p_L$, and $4.183\times10^{-6}J$ and $1.043\times10^{-6}J$ in $\W_L$, for the 80--160 and 160--320 comparisons, respectively. Successive differences decreased by a factor of $4.01$, consistent with second-order convergence. The load-activation classifications and the quoted load-work values were stable at the reported precision; in particular, $\W_L=0.166J$ at $\gamma=0.15J$ at all three step counts. The uniform fermionic-mode-loss calculation agreed with $p_L(\gamma)=e^{-\gamma t_L}p_L(0)$ to a maximum absolute population error of $2.920\times10^{-14}$ over all tested step counts, phases, and noise rates. We retain 80 steps for the reported extraction-noise curves.

\section{Baseline geometry and finite-size controls}

\begin{figure}[!htbp]
  \centering
  \includegraphics{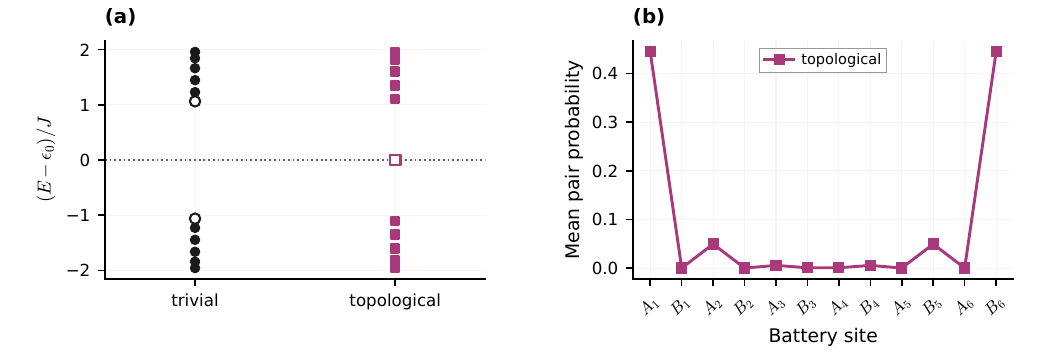}
  \caption{Clean SSH reference. Squares denote the topological phase and circles the trivial phase; open markers in (a) identify the two eigenmodes closest to midgap. (a) One-particle spectrum relative to the uniform on-site energy for the trivial and topological $N=6$ chains. The topological termination supports a near-midgap pair. (b) For the same topological $N=6$ chain, the site-resolved mean probability of the two eigenmodes closest to midgap. The finite-chain symmetric and antisymmetric pair carries equal mean weight at the two boundaries.}
  \label{figS:ssh}
\end{figure}

The clean-chain spectrum and central-pair profile establish the finite-size reference used throughout the numerical controls. Figure~\ref{figS:ssh} shows the near-midgap pair and its symmetric finite-chain edge localization for the same termination used in the charging protocol.

\begin{figure}[!htbp]
  \centering
  \includegraphics{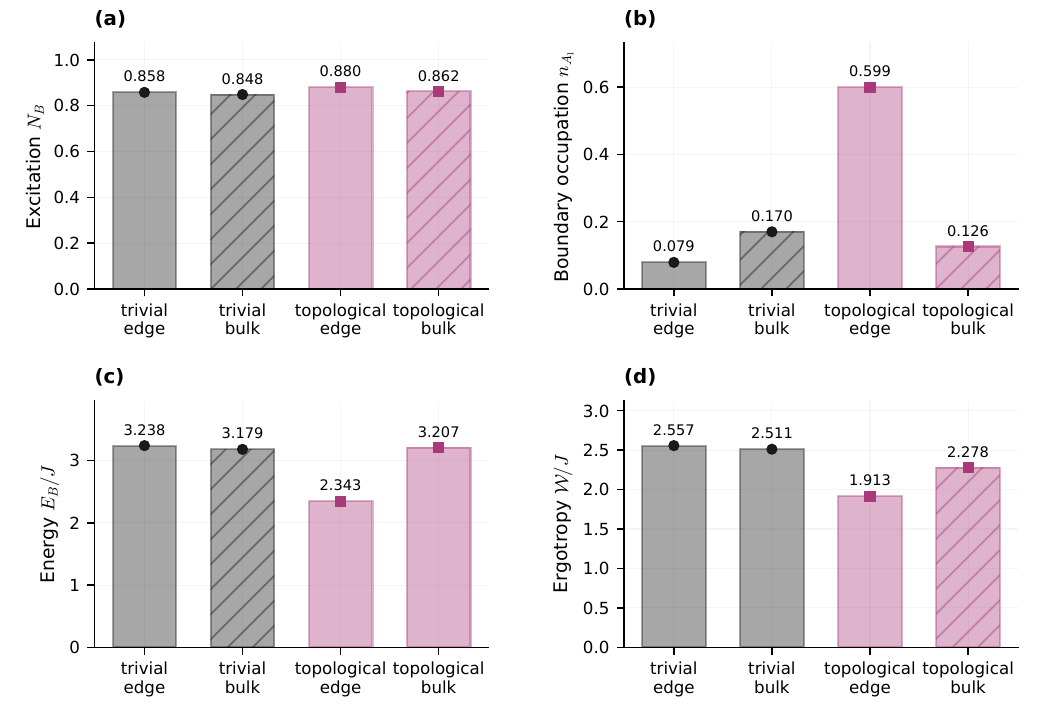}
  \caption{Full-density $N=3$ benchmark after six collisions for trivial/topological and edge/bulk geometries. Squares denote the topological phase and circles the trivial phase; hatched bars indicate bulk injection. Panels show (a) final excitation, (b) boundary occupation, (c) stored energy, and (d) global ergotropy.}
  \label{figS:four}
\end{figure}

The four-geometry benchmark in Fig.~\ref{figS:four} uses the full density matrix at $N=3$. Total excitation is similar across the four cases, while topological edge injection concentrates population at the driven boundary. The global ergotropy at baseline detuning follows a separate passive-ordering trend. The geometry control identifies the boundary eigenmode as the source of the phase contrast.

\begin{figure}[!htbp]
  \centering
  \includegraphics{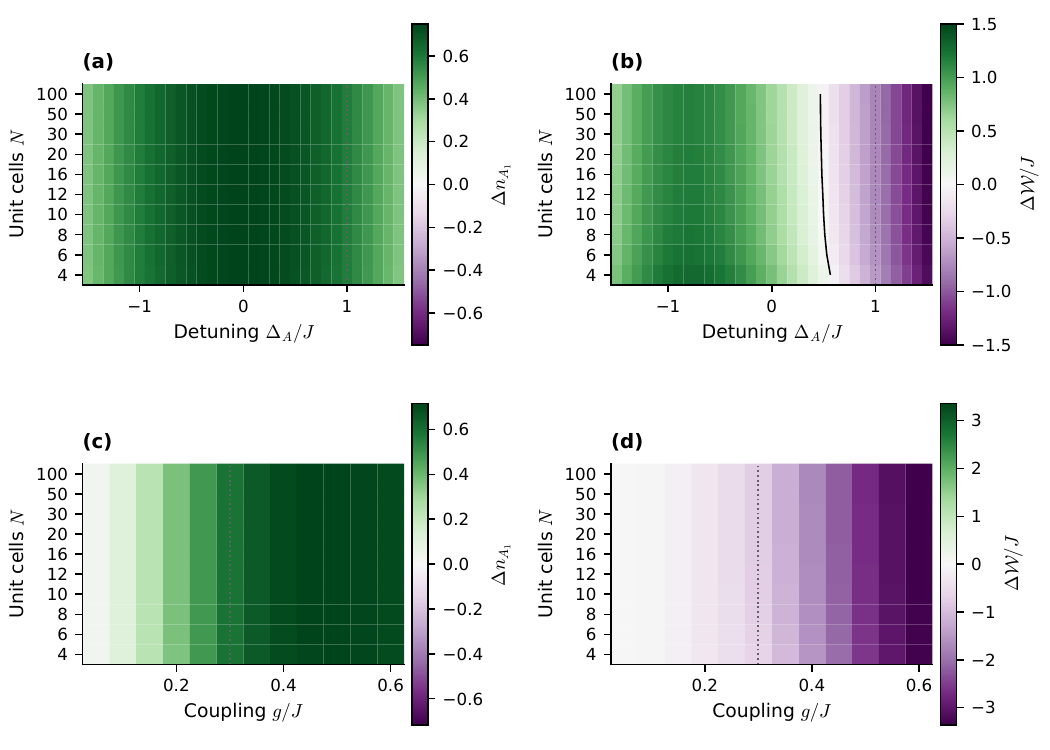}
  \caption{Fixed-control finite-size maps through $N=100$. Panels (a,b) show the topological-minus-trivial boundary-occupation and ergotropy contrasts versus ancilla detuning; panels (c,d) show the same contrasts versus charger coupling. Dotted lines mark the baseline controls, and the solid contour in (b) marks zero ergotropy contrast.}
  \label{figS:sizecontrol}
\end{figure}

\begin{figure}[!htbp]
  \centering
  \includegraphics{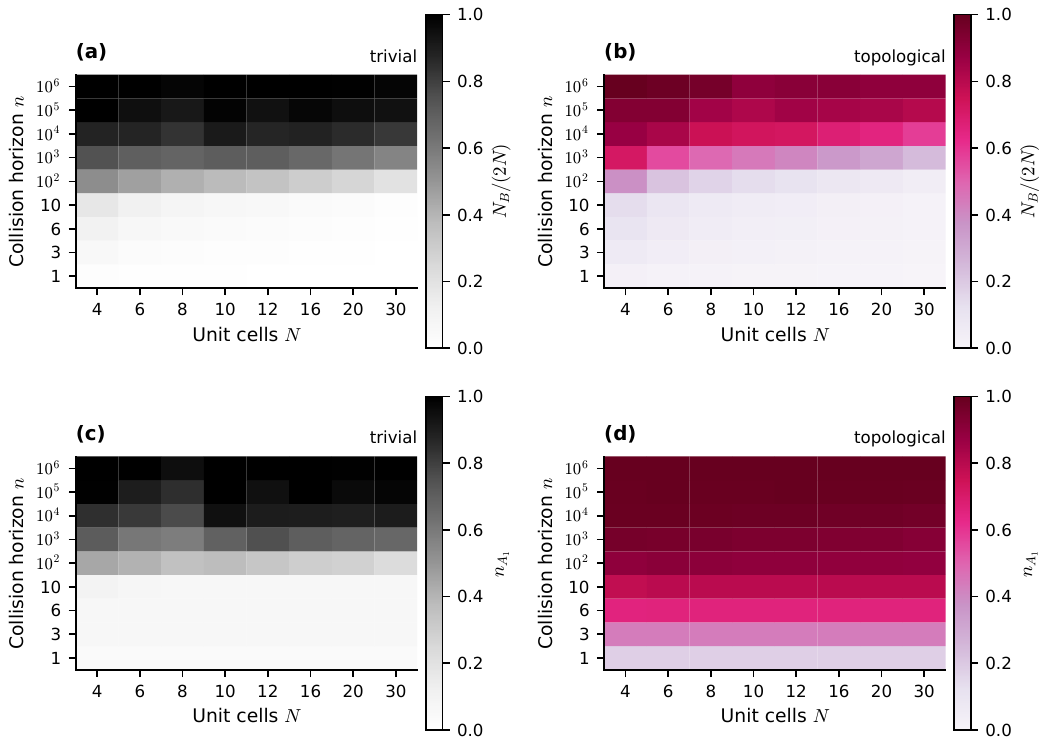}
  \caption{Collision-horizon dependence through $N=30$ and $10^6$ collisions. Panels (a,b) show the global filling fraction for the trivial and topological chains; panels (c,d) show the corresponding boundary occupation. The localized topological boundary channel fills before the remaining bulk modes reach their asymptotic state.}
  \label{figS:sizetime}
\end{figure}

At fixed controls, the topological boundary occupation is size-stable while the trivial excitation spreads. This stability follows from the boundary amplitude $\mathcal N_N=[(1-r^2)/(1-r^{2N})]^{1/2}\simeq\sqrt{1-r^2}$, which changes by less than $10^{-3}$ between $N=3$ and $N=100$ for $r=1/3$. Figure~\ref{figS:sizecontrol} shows the topological-minus-trivial boundary-occupation and ergotropy contrasts over fixed detuning and coupling controls. Figure~\ref{figS:sizetime} extends the filling comparison to long collision horizons. Positive topological--trivial contrasts occupy a finite region of the fixed control space. The long-horizon control separately shows rapid boundary filling followed by slower global saturation.

\FloatBarrier
\subsection{Fixed-control sweep across the SSH transition}

\begin{figure}[!htbp]
  \centering
  \includegraphics{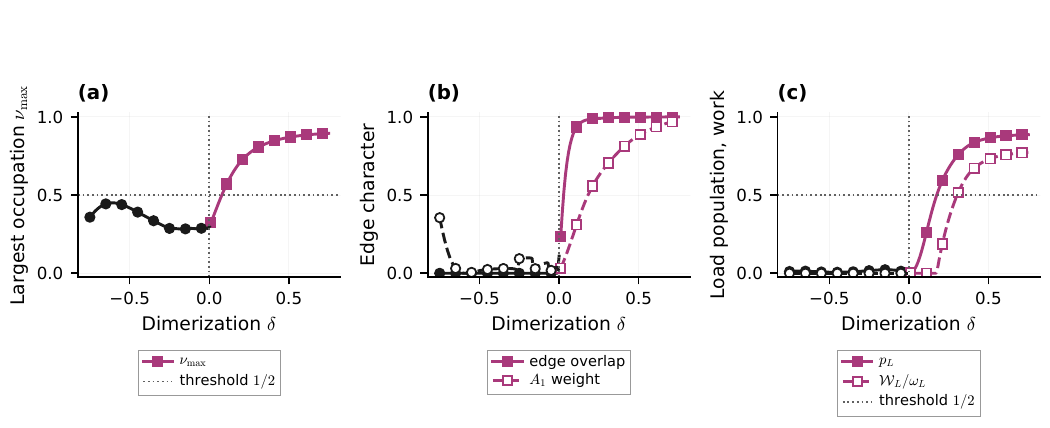}
  \caption{Full fixed-control sweep through the SSH transition for $N=100$. Squares denote $\delta>0$ (topological), and circles denote $\delta\leq0$ (trivial or critical); dotted reference lines mark $\delta=0$ and the $1/2$ thresholds. We set $v=J(1-\delta)$ and $w=J(1+\delta)$, use six resonant $A_1$ charging collisions with $g=0.3J$ and $\tau=2J^{-1}$, and apply the same empty-load pulse $(g_L,t_L)=(0.3J,5.6J^{-1})$ at every $\delta$. (a) Largest natural occupation, with $\nu_{\max}=1/2$ marking inversion. (b) Overlap of the dominant natural orbital with $\phi_{\mathrm{edge}}^{(2)}$ (solid) and its $A_1$ weight (dashed). (c) Load population (solid) and normalized load ergotropy (dashed); $p_L=1/2$ marks activation. The three diagnostics cross successively as the finite chain enters the topological regime.}
  \label{figS:transition}
\end{figure}

We parameterize the hopping pattern as $v=J(1-\delta)$ and $w=J(1+\delta)$, keeping $v+w=2J$ while crossing the bulk transition at $\delta=0$. Figure~\ref{figS:transition} reports the full $N=100$ scan with 151 points on $[-0.75,0.75]$. Six resonant collisions at $A_1$ use $\omega_A=\epsilon_0=2.5J$, $g=0.3J$, and $\tau=2J^{-1}$; the same initially empty load pulse, with $\omega_L=\epsilon_0$, $g_L=0.3J$, and $t_L=5.6J^{-1}$, is applied at every point.

For a finite open chain, let $P_2$ project onto the two one-particle eigenvectors closest to $\epsilon_0$. We define the boundary-selected reference
\begin{equation}
|\phi_{\mathrm{edge}}^{(2)}\rangle=
\frac{P_2|A_1\rangle}{\sqrt{\langle A_1|P_2|A_1\rangle}}.
\label{eqS:edge-reference}
\end{equation}
This projector definition is independent of the basis chosen within the selected two-dimensional central-energy subspace. For $\delta>0$ it approaches the finite-chain left-edge orbital; at and below the transition it provides a boundary-selected central-pair diagnostic. Finite size rounds the transition, and the response separates into three crossovers: $|\langle\phi_{\max}|\phi_{\mathrm{edge}}^{(2)}\rangle|^2=1/2$ at $\delta\simeq0.028$, $\nu_{\max}=1/2$ at $\delta\simeq0.079$, and $p_L=1/2$ for the fixed pulse at $\delta\simeq0.175$. Thus edge character develops first, an active natural orbital appears next, and the fixed boundary pulse becomes work producing after acquiring sufficient spatial and spectral overlap. The inequalities $p_L\leq\nu_{\max}$ and $0\leq\nu_j\leq1$ hold over the full sweep; the points $\delta=\mp0.5$ reproduce the two phase benchmarks.

\FloatBarrier

\section{Detuning, operating maps, and thermodynamic accounting}

\begin{figure}[!htbp]
  \centering
  \includegraphics{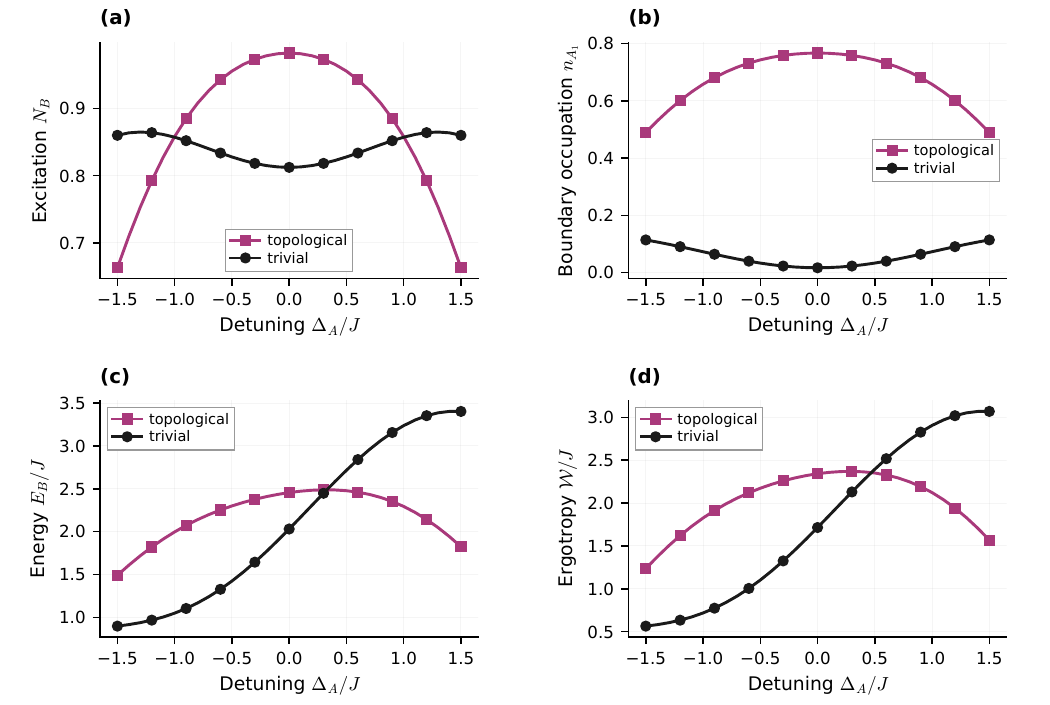}
  \caption{$N=100$ detuning scan for energy, excitation, boundary population, and global ergotropy. Squares denote the topological phase and circles the trivial phase. The topological boundary response is centered on the midgap resonance.}
  \label{figS:detuning}
\end{figure}

\begin{figure}[!htbp]
  \centering
  \includegraphics{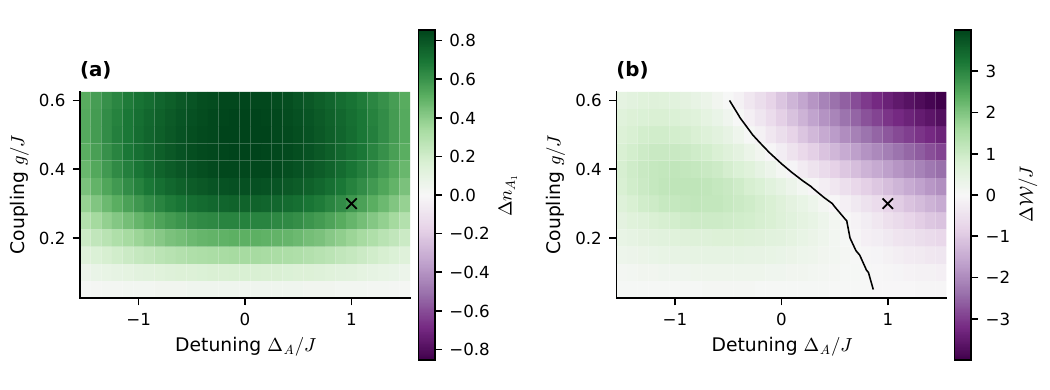}
  \caption{Joint $g$--$\Delta_A$ operating map for $N=20$ over 372 common settings in each phase. Panels show the topological-minus-trivial contrasts in (a) boundary occupation $\Delta n_{A_1}$ and (b) global ergotropy $\Delta\W$. Black crosses mark the baseline control, and the solid line in (b) marks zero ergotropy contrast.}
  \label{figS:map}
\end{figure}

The long-chain scan in Fig.~\ref{figS:detuning} distinguishes edge resonance from broadband injection. The topological response is narrow around $\Delta_A=0$, whereas the trivial chain couples to dispersive bands. The fixed-control map in Fig.~\ref{figS:map} shows where the boundary-occupation and ergotropy contrasts are positive. This spectral selectivity explains both the inverted natural occupation and the control sensitivity of the local-work advantage.

\begin{figure}[!htbp]
  \centering
  \includegraphics{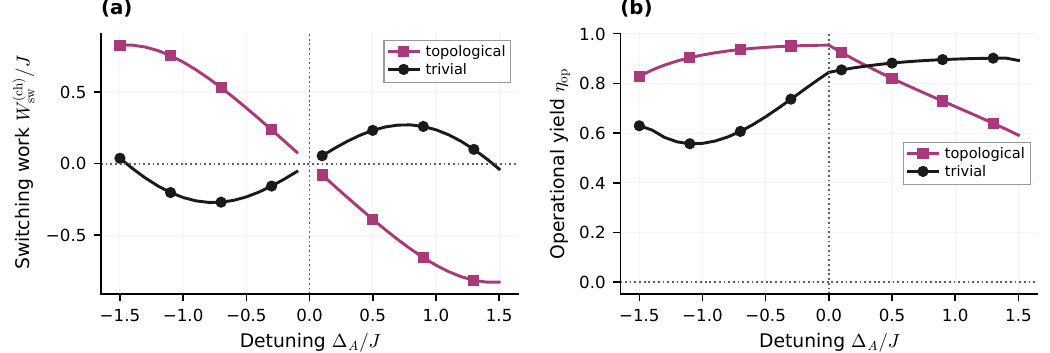}
  \caption{Charging-resource accounting at $N=100$ versus ancilla detuning. Squares denote the topological phase and circles the trivial phase. Panels show (a) aggregate charger switching work $W_{\mathrm{sw}}^{(\mathrm{ch})}$ and (b) operational ergotropic yield $\eta_{\mathrm{op}}$ for the two SSH phases. In panel (a), only the numerical off-resonant branches are plotted. The dotted vertical line marks resonance, where $W_{\mathrm{sw}}^{(\mathrm{ch})}=0$ and $E_B/N_B=\epsilon_0$ exactly in both phases by Eq.~\eqref{eqS:resonance-identities}.}
  \label{figS:thermo}
\end{figure}

The ancilla stream supplies $E_{\mathrm{in}}=\omega_AN_B$ for a vacuum battery and fresh excited chargers. Figure~\ref{figS:thermo} shows the switching contribution and the resulting yield. We report
\begin{equation}
\eta_{\mathrm{op}}=\frac{\W}{E_{\mathrm{in}}+[W_{\mathrm{sw}}^{(\mathrm{ch})}]_+}
\end{equation}
as a bounded ergotropic resource yield. Energy conservation at each separation gives $E_B=E_{\mathrm{in}}+W_{\mathrm{sw}}^{(\mathrm{ch})}$. Since $\W\leq E_B$ and $E_{\mathrm{in}}+[W_{\mathrm{sw}}^{(\mathrm{ch})}]_+\geq E_B$, one has $\eta_{\mathrm{op}}\leq1$. Likewise, $E_L\leq E_{\mathrm{in}}+[W_{\mathrm{sw}}^{(\mathrm{ch})}]_++[W_{\mathrm{sw}}^{(L)}]_+$ gives $\eta_{\mathrm{cyc}}\leq1$. The reported yield includes injected energy and positive switching work; ancilla preparation, entropy disposal, and recovery of negative switching work lie outside this accounting.

\section{Analytical edge reduction and steady state}

\begin{figure}[!htbp]
  \centering
  \includegraphics{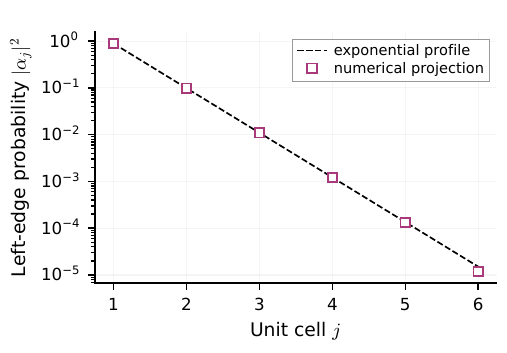}
  \caption{Analytical edge-mode reduction. The thin black dashed line is the normalized finite-chain exponential profile and open squares are the numerical midgap-subspace projection. The one-collision detuning calculation agrees with Eq.~\eqref{eqS:edge-charge}; at $\Delta_A=0$ that expression is the first step of Eq.~\eqref{eqS:edge-filling}.}
  \label{figS:edge}
\end{figure}

Figure~\ref{figS:edge} tests an analytical reduction of the resonant topological dynamics to the normalized left-edge mode. With $r=v/w$, its finite-chain boundary amplitude and the two distinct effective couplings are
\begin{equation}
\mathcal N_N=\sqrt{\frac{1-r^2}{1-r^{2N}}},\qquad
g_{A,\mathrm{eff}}=g\mathcal N_N,\qquad
g_{L,\mathrm{eff}}=g_L\mathcal N_N.
\label{eqS:edge-couplings}
\end{equation}
The one-collision edge population predicted by the charger--edge reduction is
\begin{equation}
p_{\mathrm{edge}}^{(1)}(\Delta_A)=
\frac{4g_{A,\mathrm{eff}}^2}{\Delta_A^2+4g_{A,\mathrm{eff}}^2}
\sin^2\!\left[\frac{\tau}{2}\sqrt{\Delta_A^2+4g_{A,\mathrm{eff}}^2}\right].
\label{eqS:edge-charge}
\end{equation}
At resonance a fresh excited charger updates the edge occupation according to
\begin{equation}
\nu_{\mathrm{edge}}\longmapsto
\nu_{\mathrm{edge}}\cos^2\phi+\sin^2\phi,
\qquad \phi=g_{A,\mathrm{eff}}\tau.
\end{equation}
Starting from the vacuum, $n$ collisions therefore give
\begin{equation}
\nu_{\mathrm{edge}}^{(n)}\simeq
1-\cos^{2n}\!\left(g\mathcal N_N\tau\right),
\qquad
n_{\mathrm{inv}}=\left\lceil
\frac{\ln2}{|\ln\cos^2\phi|}
\right\rceil .
\label{eqS:edge-filling}
\end{equation}
For $g=0.3J$, $\tau=2J^{-1}$, $r=1/3$, and $N=100$, Eq.~\eqref{eqS:edge-filling} gives $\nu_{\mathrm{edge}}^{(6)}=0.869$ and $n_{\mathrm{inv}}=3$. The reduction assumes that the edge coupling is small relative to the bulk gap; here $g_{A,\mathrm{eff}}\simeq0.28J$ while $|w-v|=J$. At $\Delta_A=0$, Eq.~\eqref{eqS:edge-charge} is exactly the $n=1$ term of Eq.~\eqref{eqS:edge-filling}.
After charging, an isolated edge orbital of occupation $\nu_{\mathrm{edge}}$ gives
\begin{equation}
p_L(t)\simeq\nu_{\mathrm{edge}}\sin^2(g_{L,\mathrm{eff}}t),\qquad
t_{\mathrm{swap}}=\frac{\pi}{2g_{L,\mathrm{eff}}}.
\label{eqS:edge-load}
\end{equation}
For successive fresh loads of duration $t_s$, with $\theta_s=g_{L,\mathrm{eff}}t_s$,
\begin{align}
p_{L,m}&\simeq\nu_{\mathrm{edge}}\sin^2\theta_s\cos^{2(m-1)}\theta_s,\nonumber\\
E_{L,\mathrm{cum}}^{(k)}&\simeq\omega_L\nu_{\mathrm{edge}}\left(1-\cos^{2k}\theta_s\right).
\label{eqS:edge-series}
\end{align}
For $k=6$ and $t_s=2J^{-1}$, every individual load remains passive, $p_{L,m}<1/2$, but the transferred population occupies one collective load mode with occupation $\nu_{\mathrm{edge}}(1-\cos^{2k}\theta_s)\simeq0.76$. The work thus resides in a collective load mode. A quadratic rotation among the degenerate loads costs no energy and concentrates this occupation into one load, from which $\omega_L(2p_L-1)$ is extracted.

For $r=1/3$, $N=100$, and $g_L=0.3J$, Eq.~\eqref{eqS:edge-load} gives $t_{\mathrm{swap}}=5.5536J^{-1}$, consistent with the exact grid optimum $t_L=5.6J^{-1}$. The normalization $\mathcal N_N$ changes by less than $10^{-3}$ between $N=3$ and $N=100$, so the same load duration remains near optimal over the reported sizes. Across the load-time grid displayed in the main-text local-discharge figure, the reduced model differs from the full-chain population by at most $7.26\times10^{-3}$. The residual deficit is consistent with leakage into bulk states and has the estimated perturbative scale
\begin{equation}
1-\frac{p_L}{\nu_{\mathrm{edge}}}
\sim\frac{g_L^2r^2}{(w-v)^2}.
\end{equation}
This order-of-magnitude estimate is about $1\%$ for the reported parameters, compared with the observed $0.6\%$ deficit; weaker load coupling or stronger dimerization brings the pulse closer to the bound. Finite chains also contain a right-edge partner with exponentially small splitting; the measured transfer time is short enough that this splitting is negligible at the reported sizes.

\begin{figure}[!htbp]
  \centering
  \includegraphics{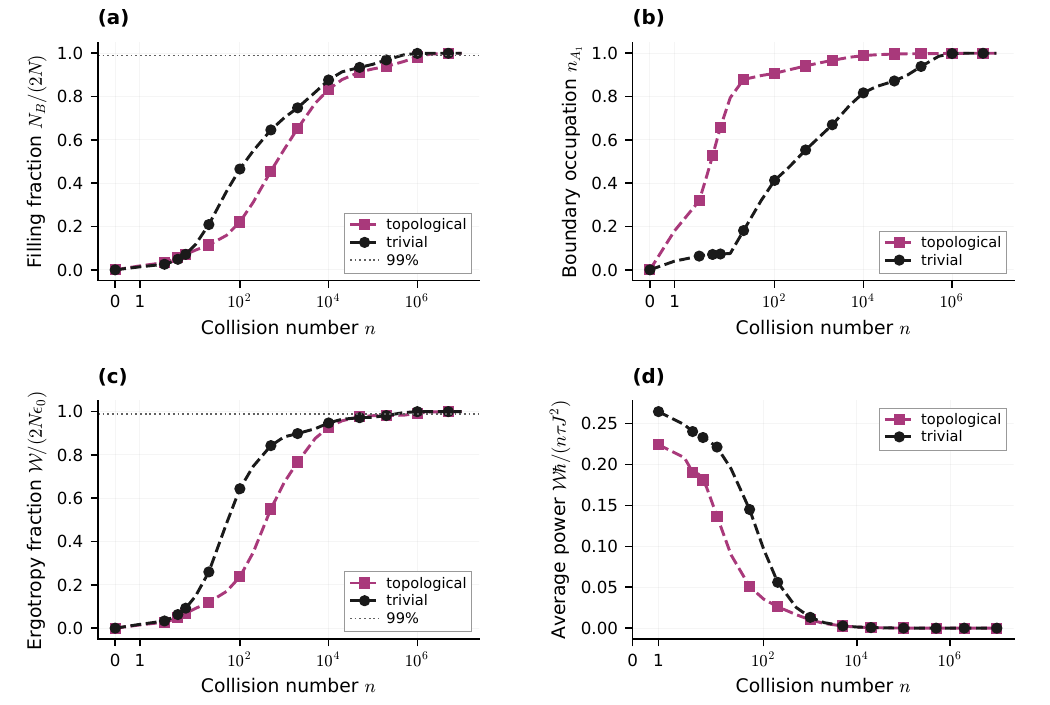}
  \caption{Collision-resolved charging and approach to the memoryless fixed point for $N=6$ through $10^7$ collisions. Squares denote the topological phase and circles the trivial phase; dashed lines use the baseline detuning $\Delta_A=J$, and dotted lines mark $99\%$ filling. The $99\%$ filling level is reached after $429{,}774$ collisions in the trivial chain and $1{,}341{,}872$ in the topological chain. The topological edge fills rapidly, while weakly addressed bulk modes set the asymptotic horizon.}
  \label{figS:steady}
\end{figure}

Repeated memoryless collisions approach the fully occupied battery, which is a fixed point of the excitation-preserving map when every fresh ancilla is excited. Figure~\ref{figS:steady} shows how rapid filling of the localized edge channel coexists with a slower global approach controlled by weakly addressed bulk modes. This explains the difference between short-horizon charging performance and the asymptotic steady state.

\FloatBarrier
\section{Selected robustness and open-system controls}

Figure~\ref{figS:control} shows the collision-control noise scan. It uses 50 realizations per point to show ensemble trends. The fixed-pulse discharge analysis in Fig.~\ref{figS:loadrobust}(d) uses 200 paired realizations at each disorder strength.

Chiral symmetry accounts for the disorder contrast. For hopping disorder that leaves the chain in the topological regime, so that the recursion below yields a normalizable profile (here $|v_j/w_j|\leq0.5$ on every bond), the semi-infinite edge mode remains pinned at $\epsilon_0$ and obeys
\begin{equation}
\alpha_{j+1}=-\frac{v_j}{w_j}\alpha_j.
\end{equation}
Hopping disorder therefore changes the boundary normalization $\mathcal N_N$ without shifting the edge energy. The chosen $t_L$ lies near the swap maximum, suppressing the first-order sensitivity of $\sin^2(g_{L,\mathrm{eff}}t_L)$ to coupling variations; at $t_L=t_{\mathrm{swap}}$, this factor changes only at second order. Hopping disorder can separately change the charged edge occupation. On-site disorder instead shifts the edge energy at first order,
\begin{equation}
\delta E_{\mathrm{edge}}\simeq
s_\epsilon\sum_j|\alpha_j|^2z_{A_j},\qquad
\sigma_E=s_\epsilon
\left[\frac{1-r^2}{3(1+r^2)}\right]^{1/2}
\simeq0.52s_\epsilon.
\label{eqS:disorder-width}
\end{equation}
Thus the long-contact centroid spread is $\sigma_E\simeq0.31J$ at $s_\epsilon=0.6J$, comparable to the standard-charge linewidth $2g_{A,\mathrm{eff}}\simeq0.57J$. This first-order detuning is consistent with the passive-load tail under strong on-site disorder.

\begin{figure}[!htbp]
  \centering
  \includegraphics{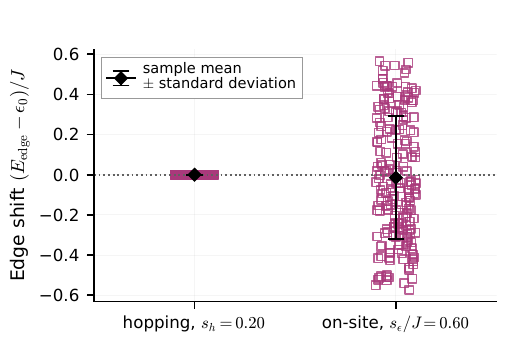}
  \caption{Realization-resolved topological edge-centroid shifts for the long-contact calculation of the main-text spectroscopy figure. The hopping-disorder ensemble uses $s_h=0.20$, and the on-site-disorder ensemble uses $s_\epsilon=0.60J$; each contains 200 realizations. Filled and open squares denote hopping and on-site disorder, respectively; black diamonds and error bars show the sample mean and one sample standard deviation.}
  \label{figS:edge-shifts}
\end{figure}

\begin{figure}[!htbp]
  \centering
  \includegraphics{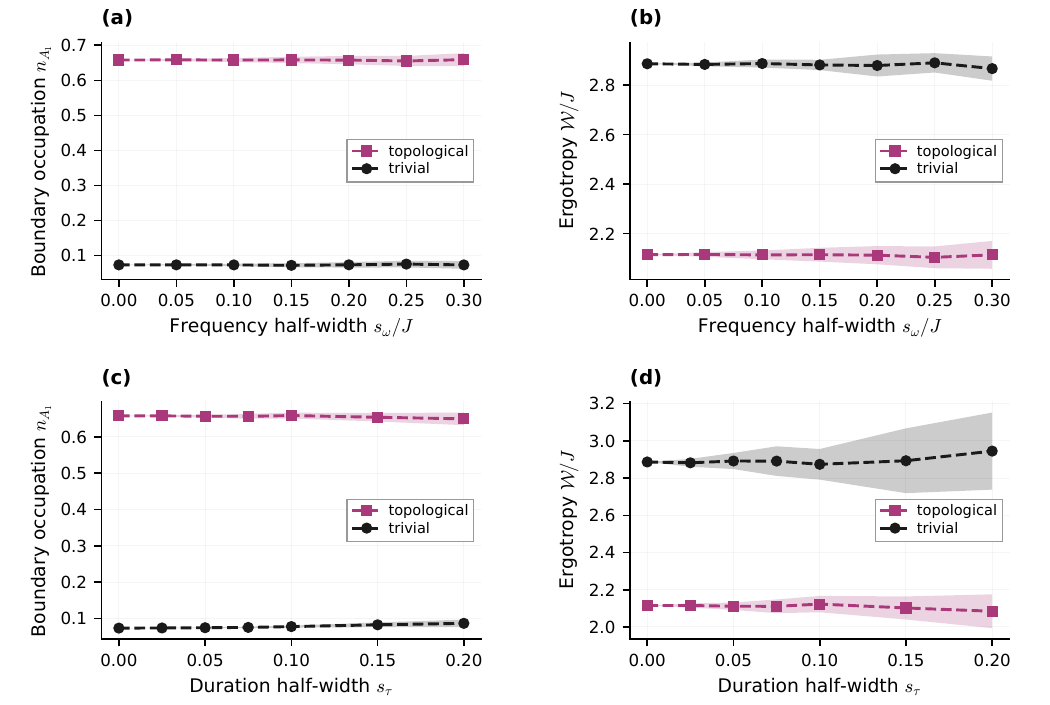}
  \caption{Collision-control noise with 50 realizations per point. Squares denote the topological phase and circles the trivial phase; dashed lines indicate the baseline detuning $\Delta_A=J$. Panels (a,b) show boundary occupation and global ergotropy versus charger-frequency half-width $s_\omega/J$; panels (c,d) show the same observables versus relative duration half-width $s_\tau$. Curves are ensemble means and bands show one sample standard deviation.}
  \label{figS:control}
\end{figure}

\begin{figure}[!htbp]
  \centering
  \includegraphics{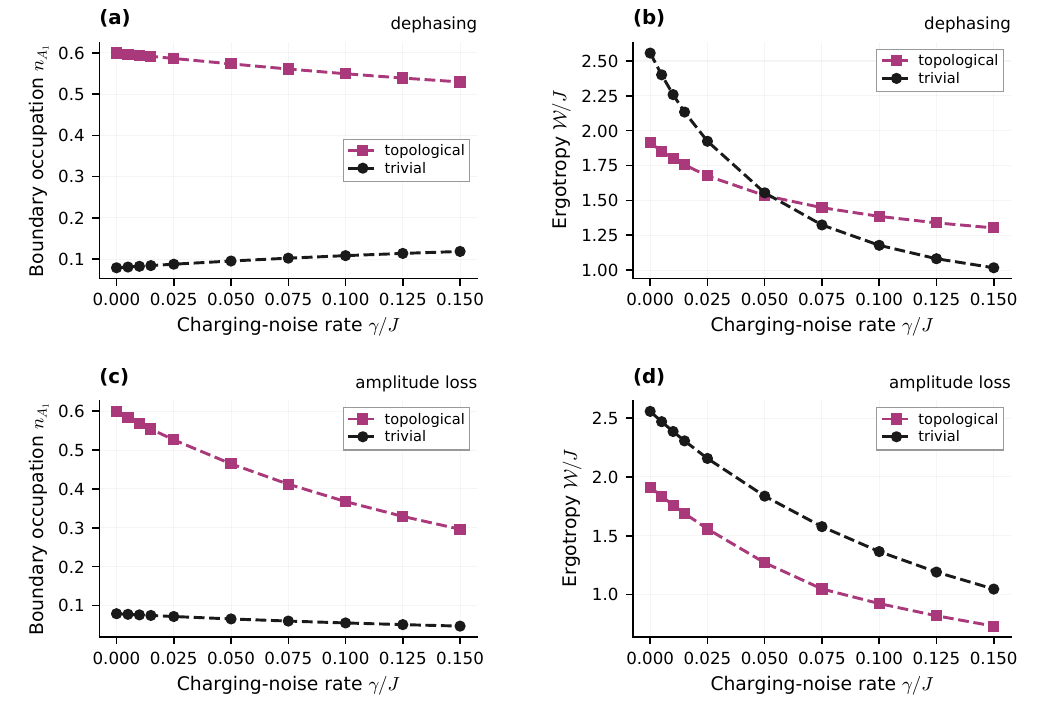}
  \caption{Full-density noisy charging at $N=3$. Squares denote the topological phase and circles the trivial phase; dashed lines indicate the baseline detuning $\Delta_A=J$. Panels (a,b) show boundary occupation and global ergotropy under local dephasing; panels (c,d) show the same quantities under local amplitude loss. Every horizontal axis uses the rate $\gamma/J$ defined in Sec.~I.}
  \label{figS:open}
\end{figure}

\begin{figure}[!htbp]
  \centering
  \includegraphics{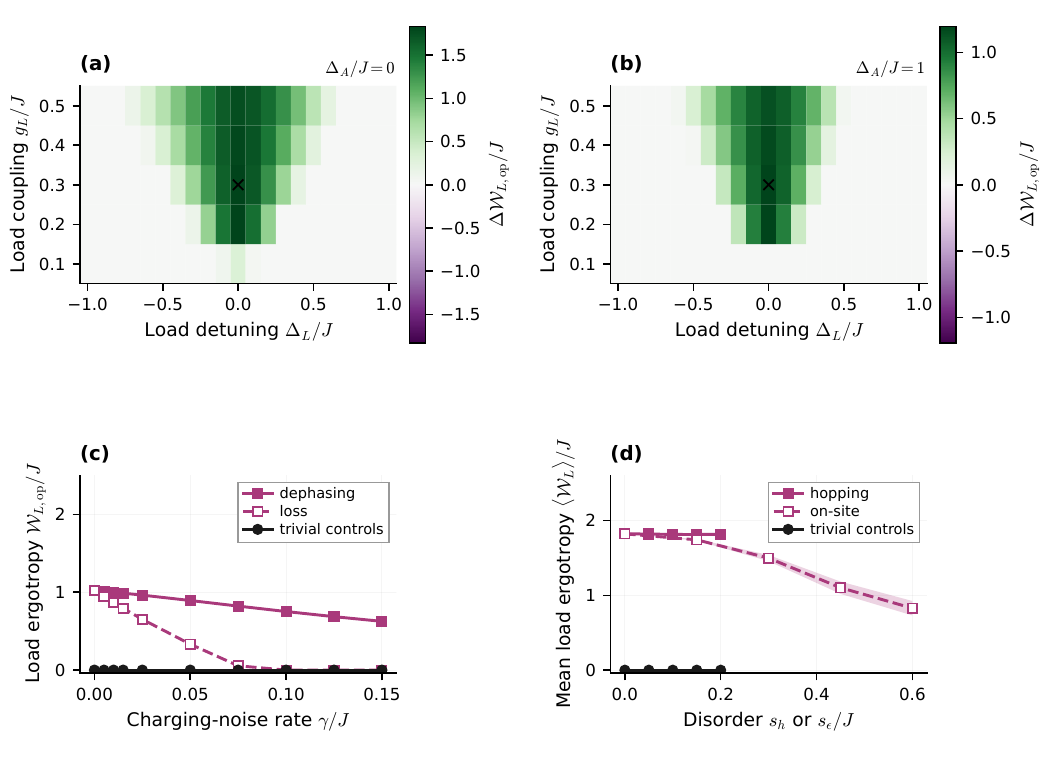}
  \caption{Robustness of fixed-pulse local work. Panels (a,b) show the phase contrast versus load detuning and coupling; crosses mark the prescribed pulse. Panel (c) shows load work after noisy full-density $N=3$ charging and ideal extraction: filled squares and solid lines denote dephasing, open squares and dashed lines denote loss, and circles denote the merged passive trivial controls. Panel (d) gives the mean over 200 static-disorder realizations with two-sided Student-$t$ 95\% confidence bands: filled squares and solid lines denote topological hopping disorder, open squares and dashed lines denote topological on-site disorder, and circles denote passive trivial controls.}
  \label{figS:loadrobust}
\end{figure}

Local qubit dephasing and amplitude loss during charging are evaluated with the full density matrix at $N=3$ in Fig.~\ref{figS:open}.

Figure~\ref{figS:loadrobust} separates noise during charging followed by ideal extraction from the extraction-stage channels reported in the main text. It also complements the main paired-contrast and active-fraction statistics with load-control maps and the mean topological load response.

For extraction-stage noise, local dephasing admits a closed evolution equation for the one-body correlation matrix. The Hamiltonian and dissipative substeps are evaluated exactly, while their combined evolution is integrated using a symmetric split-step scheme checked for timestep convergence. Uniform fermionic-mode loss commutes with the number-conserving quadratic dynamics and gives
\begin{equation}
C(t)=e^{-\gamma t}u(t)C(0)u^\dagger(t),\qquad
p_L(\gamma)=e^{-\gamma t_L}p_L(0).
\label{eqS:loss-law}
\end{equation}
The load therefore remains active for
\begin{equation}
\gamma<\gamma^*=\frac{\ln[2p_L(0)]}{t_L}\simeq0.098J.
\end{equation}
Dephasing conserves total excitation but damps the coherences that drive the swap; at $\gamma=0.15J$ the reported load work is $0.166J$. Local qubit loss coincides with fermionic-mode loss in the zero- and one-excitation sectors. Deviations are controlled by the multi-excitation weight of the state.

\FloatBarrier
\bibliography{references}